\documentclass{aa}

\usepackage{graphicx}
\usepackage{txfonts}
\usepackage[colorlinks]{hyperref}
\usepackage{color}
\definecolor{snsgreen}{rgb}{0.0, 0.620, 0.451}

\newcommand{\rev}[1]{#1}
\newcommand{\led}[1]{#1}

\hypersetup{colorlinks=true,linkcolor=blue,citecolor=blue,filecolor=blue,urlcolor=snsgreen} %

\usepackage{wasysym}
\usepackage{caption}							% to enable line breaks within captions

\usepackage{comment}						 % comments that can be switched visible/invisible
\includecomment{comment}
\specialcomment{note}{\begingroup\sffamily\color{gray}\footnotesize}{\endgroup}

\excludecomment{note}                       % switch notes off
\usepackage[disable]{todonotes}			% switches all todo notes to invisible
\usepackage{siunitx}
\DeclareSIUnit\mEarth{M_\oplus}
\DeclareSIUnit\mSun{M_\odot}
\DeclareSIUnit\rEarth{R_\oplus}
\DeclareSIUnit\year{yr}
\DeclareSIUnit\au{au}
\DeclareSIUnit\dex{dex}

\defcitealias{Emsenhuber2020}{Paper I}
\def\paperone{\citetalias{Emsenhuber2020}}
\defcitealias{Emsenhuber2020b}{Paper II}
\def\papertwo{\citetalias{Emsenhuber2020b}}

\begin{document}

   \title{The New Generation Planetary Population Synthesis (NGPPS)}
    \subtitle{III. Warm super-Earths and cold Jupiters: A weak occurrence correlation, \\ but with a strong architecture-composition link}
    \titlerunning{The New Generation Planetary Population Synthesis (NGPPS). III.}
   \author{M. Schlecker\inst{\ref{inst:mpia}}
      \and C. Mordasini\inst{\ref{inst:unibe}}
      \and A. Emsenhuber\inst{\ref{inst:ualpl},\ref{inst:unibe}}
      \and H. Klahr\inst{\ref{inst:mpia}}
      \and Th. Henning\inst{\ref{inst:mpia}}
      \and R. Burn\inst{\ref{inst:unibe}}
   \and Y. Alibert\inst{\ref{inst:unibe}}
   \and W. Benz\inst{\ref{inst:unibe}}
          }

   \institute{Max-Planck-Institut für Astronomie, Königstuhl 17, 69117 Heidelberg, Germany\\
        \email{schlecker@mpia.de} \label{inst:mpia}
        \and
        Physikalisches Institut, University of Bern, Gesellschaftsstrasse 6, 3012 Bern, Switzerland \label{inst:unibe}
        \and
        Lunar and Planetary Laboratory, University of Arizona, 1629 E. University Blvd., Tucson, AZ 85721, USA \label{inst:ualpl}
    }

    \date{Received 01 Jun 2020 / accepted 15 Aug 2020}

  \abstract
   {
   Recent observational findings have suggested a positive correlation between the occurrence rates of inner super-Earths and outer giant planets.   These results raise the question of whether this trend can be reproduced and explained by planet formation theory.}
  {Here, we investigate the properties of inner super-Earths and outer giant planets that form according to a core accretion scenario.
  \rev{We study the mutual relations between these planet species in synthetic planetary systems and compare them to the observed exoplanet population.}
  }
   {
   We invoked the Generation 3 Bern model of planet formation and evolution to simulate 1000 multi-planet systems.
   We then confronted these synthetic systems with the observed sample, taking into account the detection bias that distorts the observed demographics.
   }
   {
   The formation of warm super-Earths and cold Jupiters in the same system is enhanced compared to the individual appearances, although it is weaker than what has been proposed through observations. %
   \rev{We attribute the discrepancy to warm and dynamically active giant planets that frequently disrupt the inner systems, particularly in high-metallicity environments.
In general, a joint occurrence of the two planet types requires intermediate solid reservoirs in the originating protoplanetary disk.
   }
   Furthermore, we find differences in the volatile content of planets in different system architectures and predict that high-density super-Earths are more likely to host an outer giant.
   This correlation can be tested observationally.
   }
   {}

   \maketitle

\section{Introduction}\label{sec:introduction}

While, in the past, planet formation theories have focused on the Solar System~\citep[e.g.,][]{Pollack1996}, this focus has since shifted towards the goal of finding explanations for a whole variety of planets and planetary systems.
Important sources of constraints for these theories are the occurrence rate (or frequency) of exoplanets as a function of various orbital or physical properties as well as the fraction of stars hosting such planets~\citep[e.g.,][]{Petigura2013, Foreman-mackey2014, Hsu2018, Mulders2018a}.
In recent years, the growing sample of confirmed exoplanet systems have made such occurrence studies possible, enabling us to statistically compare theory and observations.
While the first detected exoplanet around a main-sequence star was a giant planet on a close orbit~\citep{Mayor1995}, it has now been established that ``cold Jupiters'' (CJ) in distant orbits are much more frequent but not as readily detected~\citep{Wittenmyer2020}.
Aside from spotting these types of giant planets, recent technological and methodological advances have also enabled the discovery of small, terrestrial planets, although our detection sensitivity is still limited to those \led{on} close orbits.
This development led to the discovery of an unexpected population of planets that are not present in the Solar System:
planets with masses higher than that of Earth but substantially below those of our local ice giants, that is, so-called
super-Earths~\citep[SE, e.g.,][]{Mayor2011}.
It has been estimated that they orbit \SIrange{30}{50}{\percent} of FGK stars, often in multiplanet systems~\citep{Fressin2013, Petigura2013, Zhu2018b, Mulders2018a}.

Since cold Jupiters influence their environment due to their large masses, it seems likely that they have an effect on such close-in low-mass planets~\citep[e.g.,][]{Raymond2006, Horner2010, Raymond2017}.
The open question concerns exactly how they affect the formation and subsequent evolution of inner planets and if their existence in a system facilitates the formation of super-Earths or excludes it, rather.
If hot super-Earths form in situ, there should be a positive correlation between outer giant planets and inner terrestrial systems: whenever favorable conditions enable efficient growth of planetesimals in a protoplanetary disk, both planet types can emerge~\citep{Chiang2013}.

However, in situ formation has been criticized as it is not able to account for the variety of architectures observed in these systems~\citep{Raymond2014}, thus, most current core accretion models assume orbital migration as a key ingredient~\citep[e.g.,][]{Alibert2005,Emsenhuber2020}. %
In these models, planetary cores originate from orbits that diverge from their final location through a process that typically \led{involves} inward migration.
This mechanism predicts an anti-correlation between inner super-Earths and cold giants:
due to the strong dependence of accretion timescales on the orbital radius, the innermost core is expected to grow most efficiently~\citep[e.g.,][]{Lambrechts2014a}, enabling a subsequent runaway accretion of a massive gas envelope~\citep{Pollack1996}.
The emerging giant planet now prevents cores that form further out from migrating inward to become hot super-Earths~\citep{Izidoro2015}.
On the other hand, planetary cores resulting from giant collisions can reach runaway accretion earlier, which facilitates an early growth of distant giant planets~\citep{Klahr2006}.
Models describing the growth of inner planets via pebble accretion~\citep{Ormel2010, Lambrechts2012}, which relies on a radial flux of mm to cm-sized pebbles to the inner system~\citep{Lambrechts2014a}, predict an additional impact from massive outer planets.
\rev{When} they carve a gap into the disk deep enough to generate a local pressure maximum, the inward drift of pebbles is halted \rev{just outside of the planetary orbit}~\citep{Morbidelli2012, Lambrechts2014}, possibly inhibiting the formation of inner terrestrial planets~\citep{Ormel2017, Owen2018b}.
Depending on the timing of this cut-off of pebble flux, a negative effect on the occurrence of inner super-Earths can arise.
If this scenario occurs frequently, the existence of both planet types in the same system should be rare and their occurrences anti-correlated.

However, a number of recent observational studies tested the relations between super-Earths (SE) and cold Jupiters (CJ) and found, instead, a positive correlation.
\citet{Zhu2018} measured the frequency of cold Jupiter-hosting systems in a sample of 31 systems harboring super-Earths that were first discovered by the radial velocity (RV) method.
\rev{This frequency corresponds to the conditional probability of a system harboring a cold Jupiter, given that there is at least one super-Earth in the system, P(CJ|SE).
They established P(CJ|SE) = 0.29, which is a strong enhancement compared to the fraction of field stars containing a cold Jupiter P(CJ) $\sim 0.10$.}
The opposite case, that is, the conditional probability of any super-Earth given a cold Jupiter in the system P(SE|CJ) could only be derived indirectly but was found to be even higher with the anti-correlation case definitively excluded.
\citet{Herman2019} strengthen this claim by counting five systems of transiting close-in planets in their sample of twelve long-period transiting planets.
This trend was confirmed by \citet{Bryan2019} based on a search for long-period giant companions in 65 super-Earth systems, where half of them were originally discovered by the transit method and the other half by the RV method.
Applying different boundaries for mass and period than \citet{Zhu2018}, they find P(CJ|SE) = \SI{34\pm7}{\percent} and come to the conclusion that close to all cold Jupiter-hosting systems harbor at least one super-Earth.

In light of these independent suggestions of a strong positive correlation, it is surprising that a recent RV survey that searched for super-Earth companions in giant planet-hosting systems detected none~\citep{Barbato2018}.
While their sample of 20 systems is small, their null detection is very unlikely if the correlation is indeed as high as reported.

Given this range of different results and implications for planet formation theory, great potential lies in the search for similar correlations in synthetic populations of planets produced by theoretical models.
The purpose of this study is a detailed characterization of the relations between inner super-Earths and cold Jupiters based on the core accretion theory of planet formation.
To that end, we use synthetic planetary systems that were obtained with the \rev{Generation~III} \textit{Bern} Model of planet formation and evolution \citep[hereafter \paperone]{Emsenhuber2020} to investigate the mutual influence of these planet types in and to test the observed trends.
\rev{In our simulations, we consider planetary systems around solar-type stars.}
\citet[hereafter \papertwo]{Emsenhuber2020b}, used this model to perform a population synthesis of multi-planet systems from initial conditions representative of protoplanetary disks in star forming regions.
Here, we extend that work by applying a generic detection bias and statistically compare the synthetic quantities to measured exoplanet observables.

The paper is structured as follows: In Sect. \ref{sec:methods}, we introduce our formation model with its initial conditions and describe how we prepared our synthetic data.
\rev{We present the synthetic population produced with this model, called \textit{NG76}, in Sect. \ref{sec:resultsSyn}.
Section~\ref{sec:resultsObs} pursues the comparison of our population with the observed exoplanet sample.
In Sect.~\ref{sec:discussion}, we interpret our findings and discuss their implications.
Finally, we conclude this paper by summarizing our results and predictions in Sect.~\ref{sec:conclusions}.
}

\section{Methods}\label{sec:methods}

To investigate the relations between inner rocky planets and cold gas giants, we performed a statistical comparisons between a synthetic planet population and a sample of observed exoplanets.
\rev{We focus on planetary systems around solar-type stars and fixed the stellar mass to \SI{1}{M_\odot} throughout.}
In this section, we present the global planet formation and evolution model and the choice of initial conditions used to obtain the synthetic population. Furthermore, we explain the definitions we used to classify planets and to compute their occurrence, introduce the observed sample and its biases, and demonstrate the statistical methods we applied.

\subsection{The \rev{Generation~III} Bern model}
\label{sec:BernModel}

We obtained our synthetic planetary systems using the \rev{Generation~III} \textit{Bern} global model of planetary formation and evolution (\paperone).
\rev{This semi-analytical model couples the evolution of a viscously-spreading protoplanetary disk with planet formation following the core accretion paradigm~\citep{PerriCameron1974,Mizuno1978,Mizuno1980} and a planet migration scheme (both Type~I~\citep{Paardekooper2011} and Type~II~\citep{Dittkrist2014}).
Solids are accreted via planetesimals in the oligarchic regime~\citep{IdaMakino1993,Ohtsuki2002,Thommes2003}.}
Multiple planets can form in the same disk and their mutual gravitational interaction is modeled via an N-body integrator.
The model is based on earlier work, where \citet{Alibert2005,Mordasini2009a} \rev{simulated the single embryo case}, \citet{Mordasini2012a,Mordasini2012} combined the formation phase with long-term evolution, and \citet{Alibert2013} included a formalism for concurrent formation of multiple protoplanets inside a single disk.
For a thorough description of the \rev{Generation~III} Bern Model and an outline of recent advancements of the framework, we refer to \paperone.
Additional reviews are provided in \citet{Benz2014} and \citet{Mordasini2018}.

\subsubsection{Disk model}
 The gas disk model describes the evolution of a viscous accretion disk \citep{Lust1952,Lynden-Bell1974} using an $\alpha$ parametrization as in \citet{Shakura1973} for the \rev{viscosity $\nu$}.
\rev{
We chose a viscosity parameter of $\alpha = \num{2e-3}$ that is constant throughout the disk and time-independent.
This specific value provides realistic stellar accretion rates~\citep{Mulders2017,Manara2019}.
The combination of this viscous accretion with photoevaporative mass loss results in disk lifetimes that are in agreement with observations~\citep{Haisch2001,Fedele2010,Richert2018}.
}

Following the formalism in \citet{Andrews2009}, the initial profile of the gas surface density is given by
\begin{equation}
        \label{eq:sigmaGas}
        \Sigma (r, t=0) = \Sigma_0 \left(\frac{r}{R_0}\right)^{-\beta_\mathrm{g}} \exp \left[- \left(\frac{r}{R_\mathrm{cut,g}}\right)^{2-\beta_\mathrm{g}} \right] \left(1-\sqrt{\frac{R_\mathrm{in}}{r}}\right),
\end{equation}
forming a power law with an exponential decrease at the outer edge.
Here, $\Sigma_0$ is the initial gas surface density at \rev{Jupiter's semi-major axis,} $R_0 = \SI{5.2}{au}$, \led{$\beta_\mathrm{g}$} dictates the slope of the profile, $R_\mathrm{cut,g}$ defines the location of the exponential decrease, and $R_\mathrm{in}$ marks the inner edge of the disk.
We considered only the radial dimension and assumed azimuthal symmetry.
The vertical structure was computed following the approach of \citet{Nakamoto1994}, \rev{where stellar irradiation as well as viscous heating is included to determine the midplane temperature,
\begin{equation}
        T_\mathrm{m}^4 = \frac{1}{2 \sigma} \left( \frac{3 \kappa_\mathrm{R}}{8} \Sigma(r,t) + \frac{1}{2 \kappa_\mathrm{P} \Sigma(r,t)} \right) \dot{E}_\mathrm{\nu} + T_\mathrm{irr}^4.
\end{equation}
Here, $\sigma$ is the Stefan-Boltzmann constant, $\kappa_\mathrm{R}$ is the Rosseland mean opacity, $\kappa_\mathrm{P}$ is the Planck opacity, $\dot{E}_\mathrm{\nu} = \frac{9}{4} \Sigma(r,t) \nu \Omega_\mathrm{K}^2$ is the viscous energy dissipation rate~\citep{Nakamoto1994} at Keplerian frequency $\Omega_\mathrm{K}$, and $T_\mathrm{irr}$ is the effective temperature due to stellar irradiation~\citep{HuesoGuillot2005,Fouchet2012}.
The stellar luminosity was obtained from tabulated values \citep{Baraffe2015}.
}

The surface density is further affected by gas accretion of the protoplanets as well as internal and external photoevaporation \citep{Mordasini2012a}.
\rev{We adopted inner disk boundaries} distributed like the assumed co-rotation radii of \rev{young} stars \citep{Venuti2017}.
The surface density is described by the radial power law in Equation \ref{eq:sigmaGas} with a fixed exponent $\beta_\mathrm{g} = 0.9$\rev{~\citep{Andrews2010}}.

A planetesimal disk provides material for planetary cores and evolves depending on the accretion behavior of forming protoplanets in the disk.
To account for the radial drift of particles with low to intermediate Stokes numbers\rev{~\citep[e.g.,][]{Weidenschilling1977,Birnstiel2014}, our solid disk is more compact than the gas disk.
This is imposed by a steeper slope in the planetesimal surface density as well as a decreased solid disk size.
We chose a slope index $\beta_\mathrm{s} = 1.5$, which is similar to the minimum mass solar nebula (MMSN)~\citep{Weidenschilling1977,Elbakyan2020}.}
\rev{The exponential cutoff radius, which defines the outer edge of the solid disk, was set} to half the radius of the gas disk~\citep{Ansdell2018}.

\subsubsection{Planet formation model}
Growth of the protoplanetary cores occurs via two channels: planetesimal accretion and giant impacts (collisions between protoplanets).
The planetesimal accretion rate is obtained using a particle-in-a-box approximation \citep{Safronov1969}, following the approach of \citet{Fortier2013}.
The collision probability takes into account the eccentricity and inclination distributions of the planetesimals following \citet{Ohtsuki2002}.
Although any unique planetesimal size represents a strong simplification, we assumed a uniform value of \SI{300}{\meter}.
\rev{Planetesimals of this size experience sufficient damping by gas drag to ensure viable relative velocities while exhibiting typical drift timescales that are longer than the disk lifetime~\citep{Fortier2013,Burn2019}.
They have further been shown to provide realistic accretion efficiencies to reproduce the observed population of exoplanets across various \led{planetary} mass regimes~\citep{Fortier2013}.}

The gravity of a protoplanet's core causes the attraction of gas.
Initially, the gas accretion rate of the planetary envelope is governed by the ability to radiate away the binding energy released by the accretion of both solids and gas \citep{Pollack1996}.
To determine the envelope mass and its structure, the model solves the one-dimensional internal structure equations \rev{following~\citet{Bodenheimer1986}
~\citep[also see][]{Alibert2019}.
}
During this stage, known as the "attached" phase, the boundary between planet envelope and surrounding disk is continuous.
The efficiency of cooling improves with increasing planet mass, so that the gas accretion rate increases with time.
This can eventually result in a runaway accretion of gas, where the accretion rate exceeds the amount of gas that can be supplied by the disk.
When this happens, the envelope is no longer in equilibrium with the surrounding disk and contracts \citep{Bodenheimer2000}.
In this "detached"\ phase, the internal structure equations determine the planet radius~\citep{Mordasini2012}.

\subsubsection{Orbital migration}
While embedded in a disk, planets undergo orbital migration per exchange of angular momentum with the surrounding gas \citep{Goldreich1980}.
Low-mass planets embedded in the disc migrate in the Type~I regime, while massive planets open a gap and migrate in the Type~II regime.
We consider non-isothermal Type~I migration following \citet{Paardekooper2011} with a reduction of the co-rotation torques due to the planet's eccentricity and inclination following \citet{ColemanNelson2014}.
For the Type~II migration rate, we follow \citet{Dittkrist2014} and use the fully suppressed, non-equilibrium radial velocity of the gas.
\rev{To determine the point for transition from Type~I to Type~II, we use the gap opening criterion by~\citet{Crida2006}}.
No artificial reduction factors are applied.

\subsubsection{Long-term evolution}
After the dispersal of the protoplanetary disk due to the combination of viscous accretion and photoevaporation, we modeled the thermodynamical evolution of each survived planet until a simulation time $t = \SI{10}{\giga yr}$.
The evolution module starts with a planet's internal structure at the end of the formation phase to calculate how it cools and contracts in the long term, including the effects of atmospheric escape, bloating, and stellar tides~\citep{Mordasini2012}.
\rev{Besides contraction and accretion, an additional luminosity term arises from radioactive decay of long-lived nuclides~\citep{Mordasini2012a}.}
We further took into account compositional changes of the planetary core and envelope following the method in \citet{Thiabaud2015}.

\subsubsection{N-body interactions}
Being a multi-planet simulation, the Bern model includes gravitational interaction among the growing planets during the formation phase.
We employed the \texttt{Mercury} N-body integrator \citep{Chambers1999, Chambers2012} to compute the orbital evolution of a system and detect collisions of planets.
Orbital migration as well as a damping of eccentricities and inclinations were included as additional forces.
For practical reasons, we stopped the computationally expensive N-body calculations after a simulation time of \SI{20}{\mega yr}.

\subsubsection{Monte Carlo sampling}
Reflecting the natural variation in disk properties, we performed our simulations as a Monte Carlo experiment: we repeatedly ran our model with a different set of initial conditions, which we sampled randomly from continuous distributions.
This approach enabled us to make quantifiable statistical assessments despite the complex nature of the planet formation process.

\rev{We sampled} from among four disk initial parameters: the initial gas disk mass, $M_\mathrm{gas}$, the dust-to-gas ratio, $\zeta_\mathrm{d,g}$, the mass loss rate due to photoevaporative winds, $\dot{M}_\mathrm{wind}$, and the inner disk edge, $R_\mathrm{in}$.
In addition, we randomly drew the starting locations of planetary embryos, \rev{which are instantly initialized at the beginning of the simulation}.

Constraints that can be imposed on the distribution of these parameters \led{through} observations of protoplanetary disks  are limited.
\rev{Where they were not available, we were left with theoretical arguments.
Appendix~\ref{sec:initialConditions} describes the chosen ranges for each random variable and we show their distributions in Fig.~\ref{fig:initialConditions}.}

\subsection{Synthetic planet sample}
The Monte Carlo run of our formation model yielded a population of synthetic planets that live in independent systems.
We carried out some preparatory steps before performing the statistical analysis and comparisons with the observed sample:
first, we neglected all protoplanets from further analysis that were either accreted onto the star, ejected out of the system, or did not grow beyond a total mass of \SI{0.5}{M_\oplus}.
Next, we computed the orbital period of each remaining planet from its current semi-major axis and assuming a Solar-mass host star using Kepler's Third Law.
We then categorized the sample according to the mass and period ranges in Table~\ref{tab:planetTypes} into distinct planet classes, where we considered super-Earths and cold Jupiters using the nominal definitions of \citet{Zhu2018}.
Since we are interested in the probability that a given system forms a particular planet species, we counted unique systems instead of planets to compute occurrence probabilities.

There is no general consensus about the limits in radius, mass, or composition that distinguish between different planet classes.
To facilitate comparison with observational studies, we defined planet types according to the mass and period limits in \citet{Zhu2018} and list them in Table~\ref{tab:planetTypes}.

We accounted for biases due to orbit inclinations by multiplying synthetic planet masses with  an artificial $\sin(i) $ term, where $i$ is the relative inclination between the orbital plane of the innermost planet and the line of sight to an observer.
It is reasonable to assume isotropic orientations of orbital planes, we therefore followed \citet{Mordasini2008} and drew $\sin(i)$ from the distribution,
\begin{equation}
        f(\sin(i)) = \frac{\sin(i)}{\sqrt{1 - \sin(i)^2}}.
\end{equation}

\begin{table}
\caption{Planet Classifications}             %
\label{tab:planetTypes}
\centering                          %
\begin{tabular}{c c c}        %
\hline\hline                 %
Classification & Planet Mass [$\mathrm{M}_\oplus$] & Orbital Period [d]\\\hline                        %
   super-Earth &$ \SI{2}{M_\oplus} \leq M_\mathrm{P} \sin(i) \leq \SI{20}{M_\oplus} $& $P < \SI{400}{\day}$ \\
   cold Jupiter & $M_\mathrm{P} \sin(i) \geq \SI{95}{M_\oplus}$ & $P > \SI{400}{\day}$\\
\hline
\end{tabular}
\end{table}

\rev{Our synthetic population consists of \SI{1000}{} planetary systems.
After a simulation time of \SI{5e9}{yr}, a total of \num{32030} planets on bound orbits have survived in these systems.}
Using the selection criteria in Table~\ref{tab:planetTypes}, we arrive at a sample of 538 super-Earths in 291 systems and 182 cold Jupiters in 140 systems.

\subsection{Occurrence rates and fraction of planet hosts} \label{sec:occurrenceProb}
It is crucial to distinguish between the planet occurrence rate, which constitutes a number of planets per star, and the fraction of stars hosting planets.
We consider the occurrence rate as a measure for the frequency of planets per domain in the physical parameter space and define it as
\begin{equation}
\eta = \frac{100}{N_\star}n_\mathrm{p}(\mathbf{x}),
\end{equation}
where $N_\star$ is the number of systems in the population and $n_\mathrm{p}(\mathbf{x})$ is the number of planets with properties $\mathbf{x}$ that lie in a chosen interval $d\mathbf{x}$ of the parameter space.
For the purposes of this paper, this space is spanned by combinations of orbital period, planet size, planet mass, disk solid mass, and host star metallicity.
We normalize $\eta$ to planets per 100 systems, for convenience.

We further construct the fraction of stars hosting a planet, $\mathrm{P(X)}$.
Here, $\mathrm{X}$ corresponds to a specific planet species that is defined by a parameter interval $d\mathbf{x}$.
$\mathrm{P(X)}$ is readily obtained by dividing the number of systems containing at least one planet of type $\mathrm{X}$, $N_\mathrm{X}$, by the total number of systems, that is,
\begin{equation}
\mathrm{P(X)} = \frac{N_\mathrm{X}}{N_\star}.
\end{equation}
The probability to form, for instance, a super-Earth system, $\mathrm{P(SE)}$, is then the number of unique systems containing at least one super-Earth divided by the number of systems in the population.
We note that $\mathrm{P(X)}$ is the probability that a planetary system contains at least one instance of planet species, $\mathrm{X}$, regardless of the multiplicity within a given system.
Analogously, we computed probabilities involving non-formations, $\mathrm{P(\overline{X}),}$ by counting the systems that are lacking a planet of type $\mathrm{X}$.

Conditional probabilities that quantify the fraction of systems with a planet type given that another type is present (or missing) in the system help reveal the effects of simultaneous formation of these planets in the same system.
We obtained such conditional probabilities for all possible combinations of planet types.
As an example, to compute the probability of having a cold Jupiter in a system hosting at least one super-Earth, $\mathrm{P(CJ|SE)}$, we divided the number of super-Earth systems containing a cold Jupiter by the number of super-Earth systems.
We proceeded equally with conditional probabilities of non-formations.

The uncertainties of synthetic probabilities follow a Poisson statistic since the problem is equivalent to counting measurements without errors in a binned statistic.
The requirement of independence of the individual measurements is justified since we count systems and not single planets (which could influence each other within the same system). We computed uncertainties of the conditional probabilities using Gaussian error propagation.

\subsection{Observed planet sample}
As this study investigates relations between super-Earths and cold Jupiters, we compared our synthetic population with observational samples that include these planet types.
\citet{Zhu2018} computed a variety of planet host fractions for these species and reported a positive correlation regarding their formation.
Where not stated otherwise, we refer to their numbers when using observed quantities.
A wide range of values has been reported for the fraction of stars hosting inner super-Earths P(SE), involving different mass/period limits and detection techniques~\citep[e.g.,][]{Howard2010, Howard2012, Fressin2013, Petigura2013, Zhu2018b, Mulders2018a}.
For consistency with the super-Earth definitions in \citet{Zhu2018}, we adopted $\mathrm{P(SE)}=0.30$ from \citet{Zhu2018b}.

Where quantities were missing in the literature, we obtained them using standard rules of probability theory: the observed fraction of systems that formed no super-Earth and no cold Jupiter, $P\mathrm{(\overline{SE} \cap \overline{CJ})}$, and the fraction of systems that formed super-Earths but no cold Jupiter, $P\mathrm{(SE \cap \overline{CJ})}$, were computed by applying the summation rule for probabilities.
Using the reported probability for super-Earth systems, $P\mathrm{(SE)}$, we obtain:
\begin{align}
        \mathrm{P(\overline{SE} \cap \overline{CJ})} &= 1 - \left[ \mathrm{P(SE)} + \mathrm{P(CJ)} -
        \mathrm{P(SE \cap CJ)} \right]\, \mathrm{and}\\
        \mathrm{P(SE \cap \overline{CJ})} &= \mathrm{P(SE)} - \mathrm{P(SE \cap CJ)}.
\end{align}

Taking into account that the planetary systems in our observed sample are hosted by main sequence stars and are presumably dynamically stable on $\si{\giga yr}$ timescales, we analyzed a snapshot of the synthetic planet population at \SI{5}{\giga yr}.
At this age, the protoplanetary disk has long been dispersed and the following evolutionary phase, in which thermodynamic evolution shapes the characteristics of a planet's envelope, has largely concluded~\citep{Mordasini2012a}.
It is therefore reasonable to assume that the error we introduced by assuming the same age for all stars in the sample is typically smaller than the observational uncertainties.

\subsection{Detection limit}
\label{sec:detectionLimit}
Accounting for detection limits in the observed sample, we employed a simple detection limit based on a minimum RV semi-amplitude
\begin{equation}
K = \left( \frac{2 \pi G}{P} \right)^{1/3} \frac{M_\mathrm{P} \sin i}{(M_\mathrm{P} + M_\ast)^{2/3}} \frac{1}{\sqrt{1-e^2}}
,\end{equation}
where $P$ denotes the orbital period and $e$ is the eccentricity~\citep{Cumming1999}.
\citet{Zhu2018} indicate that they removed all planet candidates with $ K < \SI{1}{\meter\per\second}$. However, their sample seems to have a sharp truncation at $K \sim \SI{2}{\meter\per\second}$ (compare their Fig. 1) which is difficult to explain by an intrinsic feature of the population. We suspect that this drop is due to a stronger detection bias than assumed and adopted a more conservative minimum $K$ of \SI{2}{\meter\per\second} for our synthetic sample to enable a more plausible comparison.

\section{Results: Synthetic population}\label{sec:resultsSyn}
\subsection{System classes}
We classified the synthetic planetary systems into four classes: systems with neither super-Earths nor cold Jupiters ($\mathrm{\overline{SE} \cap \overline{CJ}}$), systems with at least one super-Earth but no cold Jupiters ($\mathrm{SE \cap \overline{CJ}}$), systems with at least one cold Jupiter but no super-Earths ($\mathrm{\overline{SE} \cap CJ}$), and systems containing both planet types ($\mathrm{SE \cap CJ}$). Figures~\ref{fig:sysEvo_SEandCJ},~\ref{fig:sysEvo_noSEandnoCJ},~\ref{fig:sysEvo_SEandnoCJ}, and~\ref{fig:sysEvo_noSEandCJ} show the time evolution of randomly sampled systems from each of these classes. For each system, we show all planets more massive than \SI{0.5}{M_\oplus}, regardless of their detectability. Horizontal bars denote the orbital range of eccentric planets. From left to right, the columns correspond to the systems' states at simulation times \SI{0.3}{\mega yr}, \SI{1}{\mega yr}, \SI{3}{\mega yr}, the time of disk dispersal $\mathrm{t_{disk}}$, the integration time of the N-body code $t = \SI{20}{\mega yr}$, and \SI{5}{\giga yr}. At the final time, the dashed line marks an RV detection limit of $K = \SI{2}{\meter\per\second}$ and planets below this threshold are grayed out.

Overall, the systems show diverse architectures even within the same class of systems. The classes where one or both planet types are excluded often contain planets that would nominally fulfil the criteria of that planet type. These planets are either not detectable according to our chosen detection limit or were assigned an unvaforable inclination and were thus not classified as a super-Earth or cold Jupiter.

\subsection{Occurrence rates \rev{in period-radius}}

\begin{figure}
        \centering
        \includegraphics[width=\hsize]{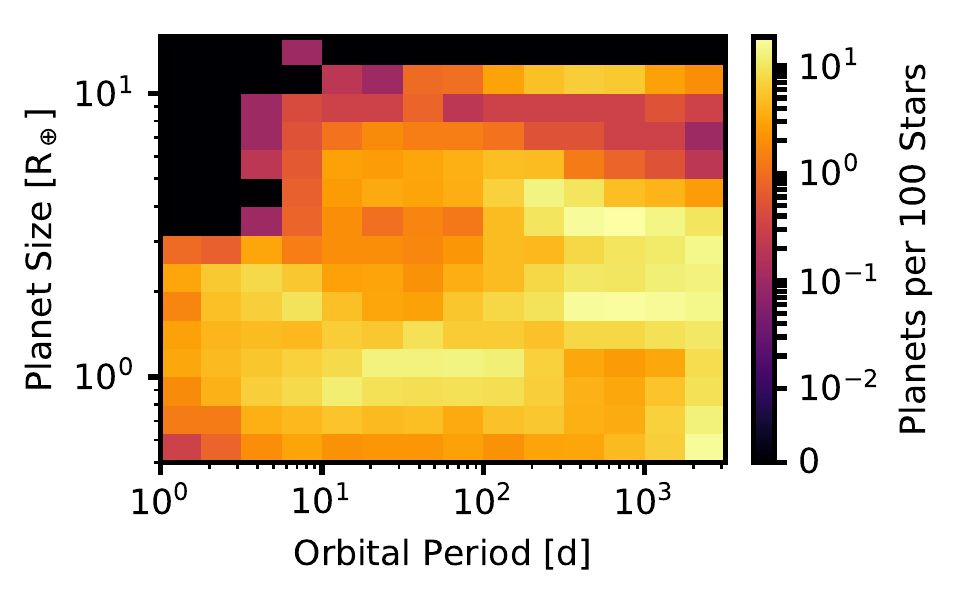}
        \caption{Occurrence map of the synthetic population. Planet occurrences are normalized to planets per 100 stars per period-radius bin, where each bin corresponds to $0.25\,\mathrm{dex}$ in period and $0.1\,\mathrm{dex}$ in planet radius, respectively. Planets with $R_\mathrm{p} < \SI{0.5}{R_\oplus}$ and beyond $P = \SI{3000}{\day}$ are not shown. Most planets are of terrestrial size and reside on intermediate orbits. A distinct group of giant planets breaks away from the remaining population at $\sim \SI{10}{R_\oplus}$.
        }
        \label{fig:occMap}
\end{figure}

Figure~\ref{fig:occMap} shows the occurrence rate, $\eta (P, R_\mathrm{p}),$ in planet radius and orbital period for the full synthetic planet population at an age of \SI{5}{\giga yr}.
Each bin covers \SI{0.25}{dex} in period and \SI{0.1}{dex} in planet radius, and their counts are normalized to planets per 100 stars.
We do not show planets beyond $P = \SI{3000}{\day}$, as our N-body integration time of \SI{20}{\mega yr} is too short to account for their large growth timescales.
We also exclude objects smaller than \SI{0.5}{R_\oplus}, which are not observable with state-of-the-art exoplanet detection techniques~\citep[e.g.,][]{Dumusque2011, Cloutier2018, Reiners2018, Bryson2019, Trifonov2020}.
The majority of planets are of terrestrial size and reside on intermediate or wide orbits. A sub-population of Jupiter-sized planets ($R_\mathrm{P} \sim \SI{11}{R_\oplus} $) is clearly differentiated from the contiguous remaining population and preferentially populates the period range of a few hundred to $\sim \SI{1000}{\day}$.

\subsection{\rev{Occurrence rates in period-mass}}
\begin{figure}
        \centering
        \includegraphics[width=\hsize]{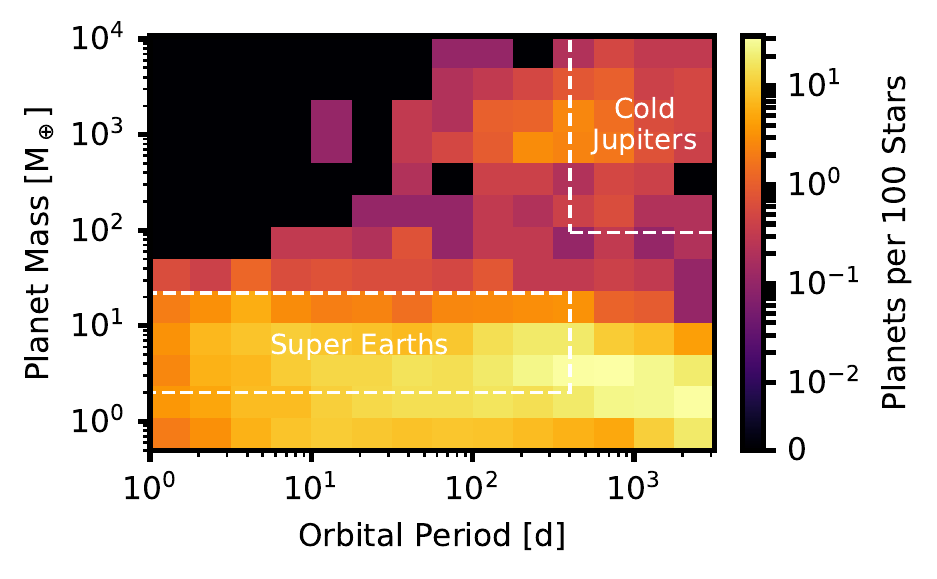}
        \caption{Occurrences of the synthetic population in the mass-period plane. Normalization and binning are the same as in Fig.~\ref{fig:occMap}, except that each planet mass bin corresponds to $1/3 \,\mathrm{dex}$. White lines border the mass-period limits for super-Earths and cold Jupiters, respectively.
        }
        \label{fig:occMapMasses}
\end{figure}

Similarly to Fig.~\ref{fig:occMap}, Fig.~\ref{fig:occMapMasses} shows synthetic planet occurrence rates in the mass-period plane $\eta (P, M_\mathrm{P})$.
The occurrence is normalized to number of planets per 100 stars per mass-period interval.
Following the same argument as for very small radii, we refrain from considering planets less massive than \SI{0.5}{M_\oplus}.
As in Fig.~\ref{fig:occMap}, we exclude planets on periods beyond \SI{3000}{\day}.
Unsurprisingly, the distribution is similar to the radius-period diagram with high-mass planets more dispersed, although they still form a distinct population.
There are no distinct populations of hot and cold Jupiters, only a small number of giants with $P \sim \SI{10}{\day}$ separates from the main group of giant planets at intermediate to large orbital distances.
The latter is only partly included in our definition of cold Jupiters owing to the comparability with the observed planet sample.
Rocky planets of terrestrial to super-terrestrial mass occupy predominantly periods of hundreds to thousands of days; the bulk of planets populating outer regions falls outside our nominal super-Earth definition.
We note that compared to previous population syntheses that lacked intermediate-mass inner planets~\citep[e.g.,][]{Mordasini2009a}, our current model produces a significant number of super-Earths:  291 out of 1000 systems harbor a planet that obeys our criteria for a super-Earth (compare Table~\ref{tab:planetTypes}). This difference is mainly caused by our improved description of planet migration, which in particular treats the shift from type I to type II migration self-consistently. No artificial inhibition factors for type I migration %
are necessary to reproduce observed period distributions.

\subsection{Relation to disk properties}\label{sec:diskProps}
\begin{figure*}
        \centering
        \includegraphics[width=\hsize]{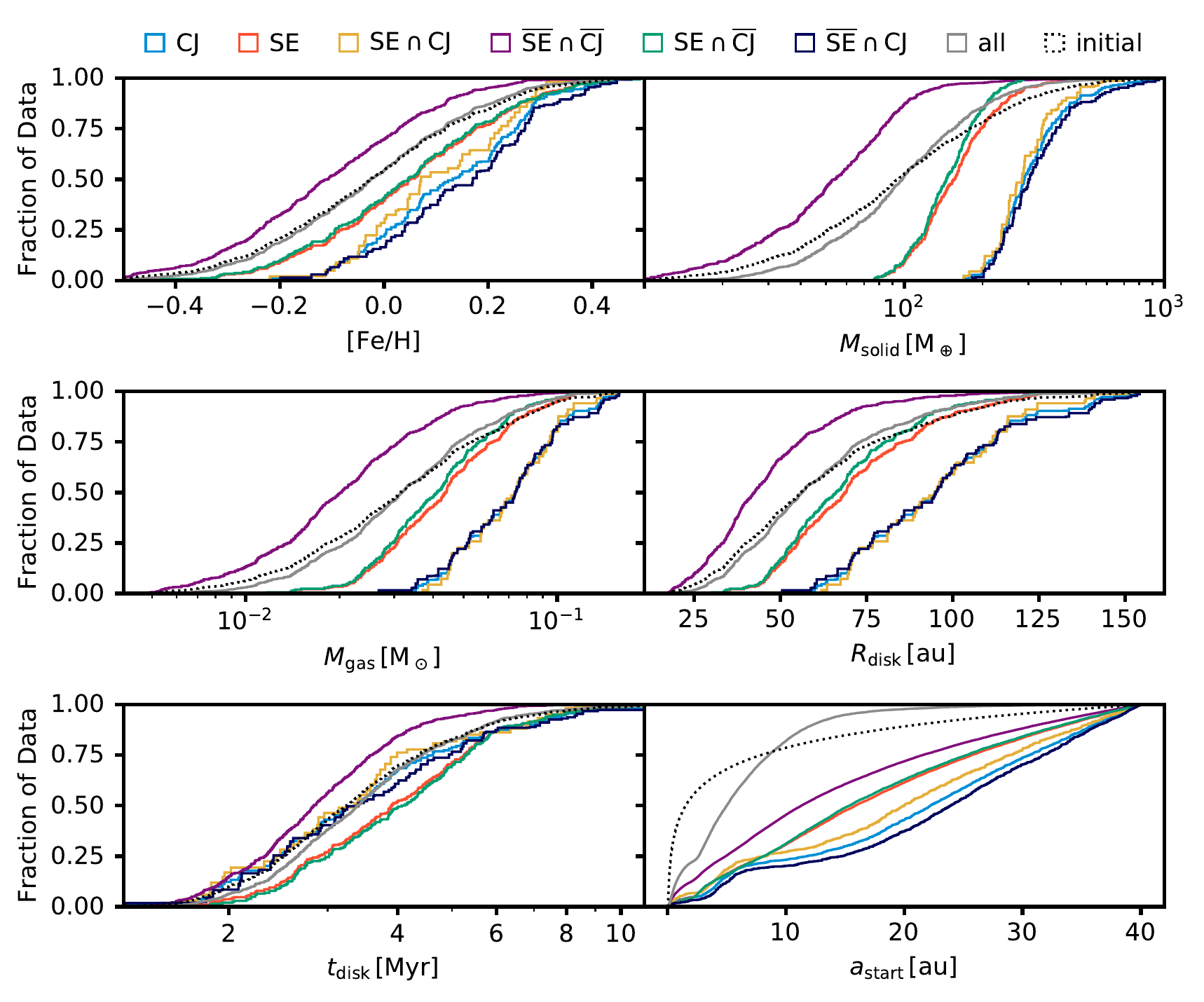}
        \caption{Initial conditions for different populations. For each parameter, we show empirical distribution functions for all combinations of SE and CJ, plus for the entire population of survived planets with $M_\mathrm{P} > \SI{0.5}{M_\oplus}$. The dotted lines show the initial distributions for the simulations. The CJ-hosting populations form compact clusters in most parameters, whereas the SE populations spread more depending on the existence of CJ in the systems.}
        \label{fig:initialConditionsOfPopulations}
\end{figure*}

In contrast to studies that focus on observed planet populations, for the synthesized population, we have the full history of each simulated system at hands.
This includes the initial properties and evolution of the protoplanetary disk in which the synthetic planets formed (or not). Figure~\ref{fig:initialConditionsOfPopulations} reveals the distributions of these features for each of the system classes in Table~\ref{tab:probs} as well as for all planets with masses of $> \SI{0.5}{M_\oplus}$ (``all'', gray lines).
In each case, we show the parameter distributions of all planets that survived the entire formation and evolution phase, for example, the ``SE'' population contains all surviving planets in super-Earths-hosting systems and not only their super-Earths.
For comparison, the dotted lines (``initial'') denote the distributions for the complete set of simulations.
We note that the gas disk radius $R_\mathrm{cut,g}$ is not an independent parameter but a unique function of the gas disk mass $M_\mathrm{gas}$.
Table~\ref{tab:initialConditionsOfPopulations} contains the 16th, 50th, and 84th percentiles of these quantities for each system class.

In all physical disk parameters related to available planet material (metallicity, solid and gas mass, and disk size), the same three distinct populations are differentiated: systems without super-Earths or cold Jupiters, systems that formed intermediate-mass planets, such as super-Earths, and systems that formed cold Jupiters.
This clustering is particularly illustrative in $M_\mathrm{solid}$ when retraced from low to high values:
\begin{itemize}
        \item $\mathrm{\overline{SE} \cap \overline{CJ}}$: at low solid masses of tens of \SI{}{M_\oplus}, only low-mass planets occur that do not reach super-Earth mass or beyond.
        \item ``All'' class represents all survived planets and thus closely resembles the initial conditions.
        \item $\mathrm{SE}$ and $\mathrm{SE \cap \overline{CJ}}$: in disks of intermediate supplies of solids, cores of several \SI{}{M_\oplus} can form which result in super-Earths.
        \item $\mathrm{SE \cap CJ}$, $\mathrm{CJ}$, and $\mathrm{\overline{SE} \cap CJ}$: from $M_\mathrm{solid} \gtrsim \SI{200}{M_\oplus}$, cold Jupiters can form. As shown below, these giant planets can pose a threat to inner super-Earth systems, which are frequently destroyed in $\mathrm{\overline{SE} \cap CJ}$ systems.
\end{itemize}

The starting position of the planetary embryo is a particularly decisive feature (compare Schlecker et al., in prep.) and shows a separate pattern: while the overall population, which is dominated by terrestrial-mass planets, is shifted to small orbits, $< \SI{10}{au}$, all other populations follow a more balanced distribution.
Again, the cold Jupiter-hosting populations are differentiated.
They show a bimodal distribution of initial orbits that divides them into planets we labeled as super-Earths or cold Jupiters and companions in the same systems that are undetectable.

The disk lifetime is rather insensitive to the outcome, but shows a similar clustering of system classes as the other parameters.
While $\mathrm{\overline{SE} \cap \overline{CJ}}$ systems, which consist largely of low-mass planets, have a median disk lifetime of \SI{2.8}{\mega yr}, the disks of systems hosting super-Earths but no cold Jupiters last for \SI{4.1}{\mega yr} on average.
For a more detailed analysis of the links between disk properties and resulting planet types, see Schlecker et al. (in prep.) and Mordasini et al. (in prep.).

\begin{figure}
        \centering
        \includegraphics[width=\hsize]{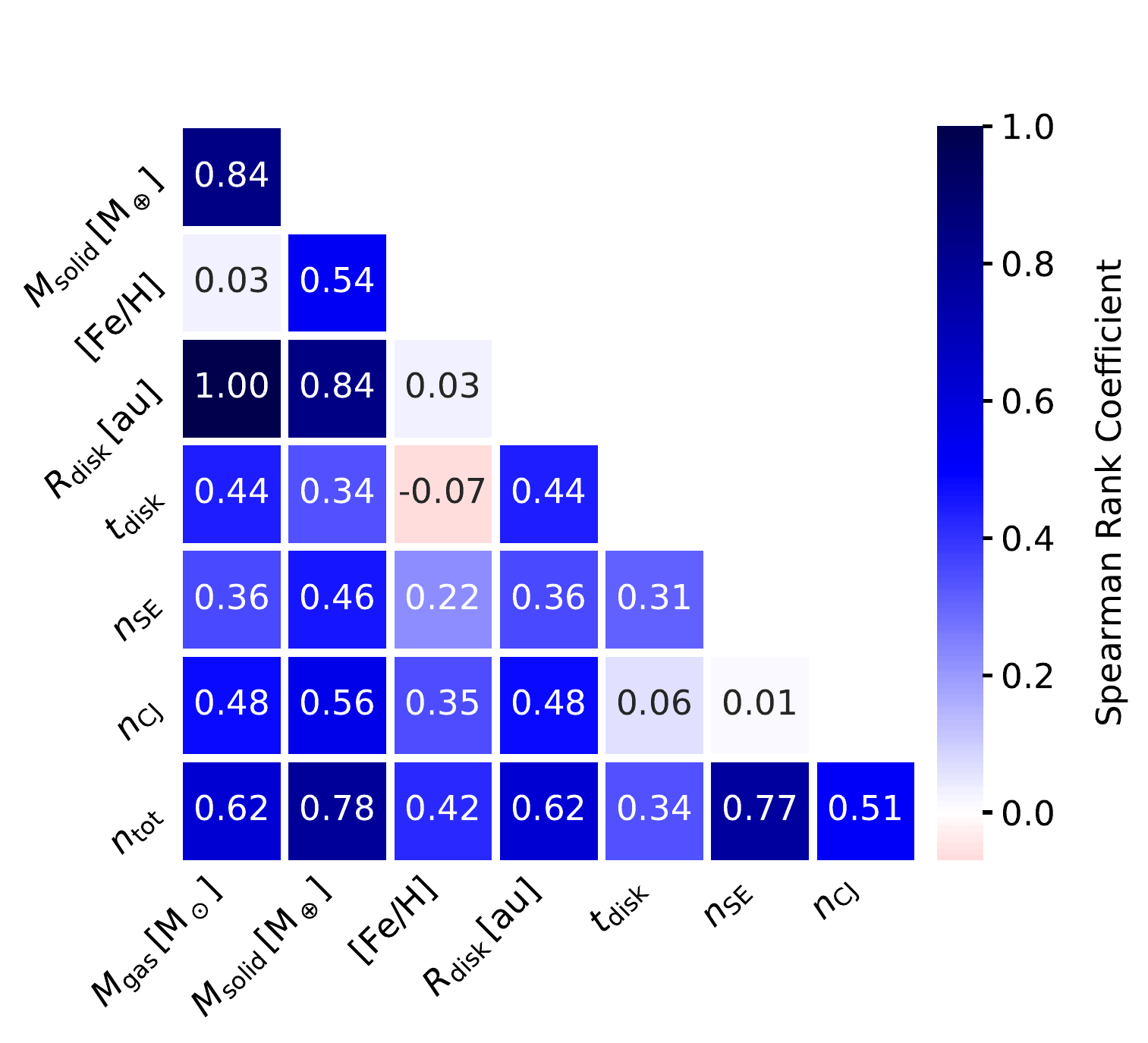}
        \caption{Correlation Map of disk properties and planet occurrence rates.          For every combination of two quantities, we compute the Spearman rank coefficient to assess mutual correlations. The (identical) upper triangle and the self-correlating diagonal are removed for clarity. Almost all parameters show positive correlations and of all disk features, the solid material supply of the disk shows the highest correlation with all planet occurrences.
                }
        \label{fig:correlationMap}
\end{figure}

We now look at correlations between disk initial conditions and planet occurrence on the system level.
To do so, we computed Spearman's rank correlation coefficient $\rho$~\citep{Spearman1904} for combinations of disk parameters (compare Table~\ref{tab:modelParams}) and occurrences of a planet type.
The coefficient ranges from $-1$ to $+1$, where a positive (negative) coefficient denotes a positive (negative) rank correlation between two variables and $\rho = 0 $ corresponds to no correlation.
All synthetic systems were included.
The correlation map in Fig.~\ref{fig:correlationMap} includes the initial gas disk mass, $M_\mathrm{gas}$, the initial solid disk mass, $M_\mathrm{solid}$, the host star metallicity, [Fe/H], the exponential cutoff radius of the gas disk, $R_\mathrm{cut,g}$, and the disk dispersal time, $t_\mathrm{disk}$. %
It further incorporates occurrence rates for three different planet types: $n_\mathrm{SE}$ and $n_\mathrm{CJ}$ are the number of super-Earths and cold Jupiters per system, respectively. \led{$n_\mathrm{tot}$} is the per-system frequency of all planets more massive than \SI{0.5}{M_\oplus}.
All occurrence rates show positive correlations with $M_\mathrm{gas}$, $M_\mathrm{solid}$, and $R_\mathrm{cut,g}$ (which is itself a function of $M_\mathrm{gas}$). \led{
$n_\mathrm{CJ}$} and $n_\mathrm{tot}$ are also moderately correlated with metallicity, and $n_\mathrm{SE}$ and $n_\mathrm{tot}$ show some dependence on $t_\mathrm{disk}$.
For all occurrence rates, the strongest correlation is obtained with $M_\mathrm{solid}$.

\begin{figure}
        \centering
        \includegraphics[width=\hsize]{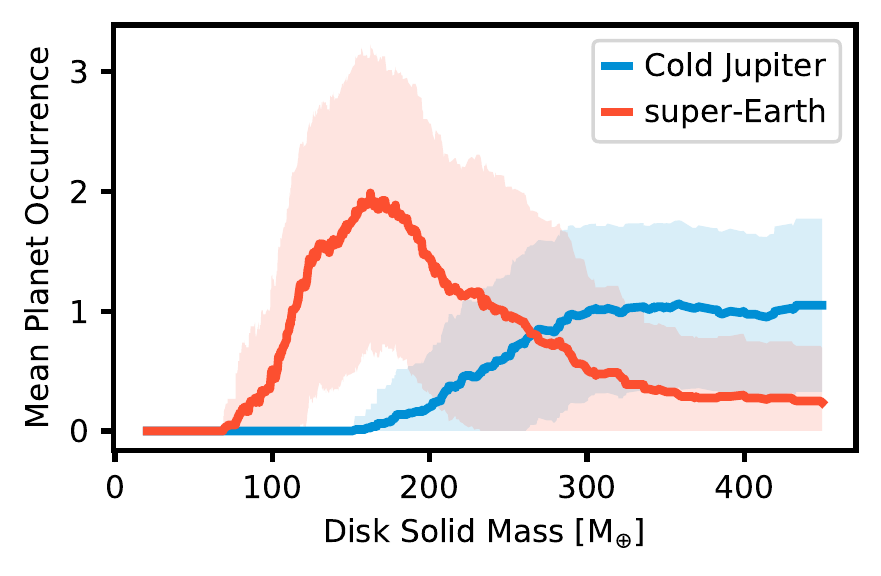}
        \caption{Mean planet occurrence per system as a function of initial solid disk mass. A rolling mean occurrence along the solid disk mass axis is shown for each planet type separately. Shaded areas cover $\pm 1$ standard  deviation of the rolling mean. Cold Jupiters form only in disks exceeding $M_\mathrm{solid} \approx \SI{150}{M_\oplus}$, where their occurrence shows a shallow positive correlation with solid mass. For super-Earths, there is a sharp increase to a maximum occurrence at $\sim\SI{160}{M_\oplus}$, then it drops before flattening out at $\sim\SI{250}{M_\oplus}$.
                }
        \label{fig:occurrenceSolidDiskMass}
\end{figure}

In order to identify trends of occurrence rates of the different planet types with this parameter, we compute a rolling mean along the solid mass axis and plot the corresponding mean planet occurrences of cold Jupiters and super-Earths, respectively (Fig.~\ref{fig:occurrenceSolidDiskMass}). The rolling window moves with step size one and consists of 80 systems; this window size is a trade-off between resolution and robustness against random variations. Shaded areas in the plot cover one standard deviation around the mean. For cold Jupiters, we obtain a monotonically increasing mean occurrence rate that starts around \SI{150}{M_\oplus} and flattens out arriving at $\sim1.0\pm0.6$ planets for $M_\mathrm{solid}\gtrapprox \SI{300}{M_\oplus}$.

The picture is very different for super-Earths. Their formation starts at lower disk masses with a steep increase in frequency up to a peak of $1.9\pm1.2$ planets per system at $\SI{160}{M_\oplus}$ in solids. At higher disk masses, the occurrence drops below unity and slightly descends beyond $\SI{300}{M_\oplus}$. Interestingly, these strong variations coincide with features in the cold Jupiter occurrence.
This points to the destruction of inner rocky planetary systems by emerging outer giants in systems with very high initial solid disk mass $M_\mathrm{solid}$.
Generally speaking, systems harboring both super-Earths and cold Jupiters require disks with intermediate reservoirs of solids.
If they are too small, no giant planets form.
If they are too large, the super-Earths are destroyed and only giants remain.

\section{Results: comparison with observations}\label{sec:resultsObs}
In this section, we aim to compare the planet population \textit{NG76} produced by our formation model with observed exoplanets.
We focus on the populations of super-Earths and cold Jupiters and compare them to recent results based on data from ground-based radial velocity measurements and from the \textit{Kepler} mission~\citep{Borucki2010a}.
For a confrontation of observed and calculated planetary bulk densities in Sect.~\ref{sec:iceMassFractions}, we compiled our own sample of confirmed exoplanets .

\subsection{Fractions of planet hosts}
To understand the relations between close-in super-Earths and outer gas giants and to constrain their mutual influence in the formation of planetary systems, we are interested in the fractions of planetary systems that form (and maintain) these planets.

\begin{table*}
        \caption{\label{tab:probs}Fractions of Stars hosting super-Earths and cold Jupiters.}
        \centering

        \begin{tabular}{lccccccc}
                \hline\hline
                Probability& Observed\tablefootmark{a}& Full Population\tablefootmark{e} & [Fe/H] < -0.2 & -0.2 < [Fe/H] < 0.2 &  0.2 < [Fe/H] & $\bar{\mu}$ & $\bar{e}$\\
\hline
$\mathrm{P(SE)}$ & 0.30\tablefootmark{b} &                                  $0.29\pm$0.02 & 0.13$\pm$0.02 &  0.31$\pm$0.02 & 0.43$\pm$0.05 & 1.8$\pm$0.9 & 0.08$\pm$0.10\\
$\mathrm{P(CJ)}$ & 0.11\tablefootmark{d} &                                          $0.14\pm$0.01 &  <0.01        &  0.12$\pm$0.01 & 0.42$\pm$0.05 & 1.3$\pm$0.5 & 0.18$\pm$0.19\\
$\mathrm{P(SE \cap CJ)}$ & 0.09 &                                                   $0.05\pm$0.01 &  <0.01        &  0.05$\pm$0.01 & 0.10$\pm$0.03 & 3.7$\pm$0.9 & 0.11$\pm$0.13\\
$\mathrm{P(\overline{SE} \cap \overline{CJ})}$ &0.69\tablefootmark{b} &    $0.62\pm$0.02 & 0.87$\pm$0.07 &  0.62$\pm$0.03 & 0.25$\pm$0.04 & 1.8$\pm$0.8 & 0.09$\pm$0.10\\
$\mathrm{P(SE \cap \overline{CJ})}$ & 0.21\tablefootmark{b} &               $0.24\pm$0.02 & 0.12$\pm$0.02 &  0.26$\pm$0.02 & 0.32$\pm$0.05 & 2.2$\pm$1.0 & 0.06$\pm$0.10\\
$\mathrm{P(\overline{SE} \cap CJ)}$ & 0.01\tablefootmark{c} &                       $0.09\pm$0.01 &  <0.01          &  0.07$\pm$0.01 & 0.32$\pm$0.05 & 2.1$\pm$0.9 & 0.19$\pm$0.18\\
\hline
$\mathrm{P(SE|CJ)}$ & 0.90$\pm$0.20 &                                               $0.34\pm$0.06 &  1.0$\pm$1.41 &  0.41$\pm$0.09 & 0.25$\pm$0.07 & --          & --\\
$\mathrm{P(CJ|SE)}$ & 0.32$\pm$0.08 &                                               $0.16\pm$0.03 & 0.04$\pm$0.04 &  0.15$\pm$0.03 & 0.24$\pm$0.07 & --          & --\\
$\mathrm{P(\overline{SE}|CJ)}$ & 0.10$\pm$0.20 &                                    $0.66\pm$0.09 &  <0.01        &  0.59$\pm$0.11 & 0.75$\pm$0.14 & --          & --\\
$\mathrm{P(\overline{CJ}|SE)}$ & 0.68$\pm$0.08 &                                    $0.84\pm$0.07 & 0.96$\pm$0.27 &  0.85$\pm$0.09 & 0.76$\pm$0.14 & --          & --\\
\hline
        \end{tabular}

        \tablefoot{The top part shows the fraction of stars harboring (lacking) super-Earths (SE), cold Jupiters (CJ), and combinations thereof \rev{at an age of \SI{5}{\giga yr}}.
        The bottom panel shows conditional probabilities $\mathrm{P(A|B)}$ where $A$ denotes the existence of at least one instance of planet type $A$ in a given system and $\overline{A}$ denotes its non-existence. Uncertainties of probabilities are based on Poisson errors.
        The last two columns list the mean planet multiplicity $\bar{\mu}$ and mean eccentricity $\bar{e}$ with their standard deviations.
        While for P(SE) (P(CJ)), this takes into account only super-Earths (cold Jupiters); for the other classes we consider all planets with $K > \SI{2}{\meter\per\second}$.\\
                \tablefoottext{a}{if not stated otherwise, probabilities are from \citet{Zhu2018} using their ``nominal'' super-Earth definition where $\mathrm{M_P} \sin i < \SI{20}{M_\oplus} $.}\\
                \tablefoottext{b}{quoting for P(SE) the fraction of \textit{Kepler} systems hosting super-Earths in \citet{Zhu2018b}}\\
                \tablefoottext{c}{order of magnitude estimate by \citet{Zhu2018}}\\
                \tablefoottext{d}{as estimated by \citet{Cumming2008}}\\
        \tablefoottext{e}{population \textit{NG76} in the NGPPS series}
        }
\end{table*}

\begin{figure}
        \centering
        \includegraphics[width=\hsize]{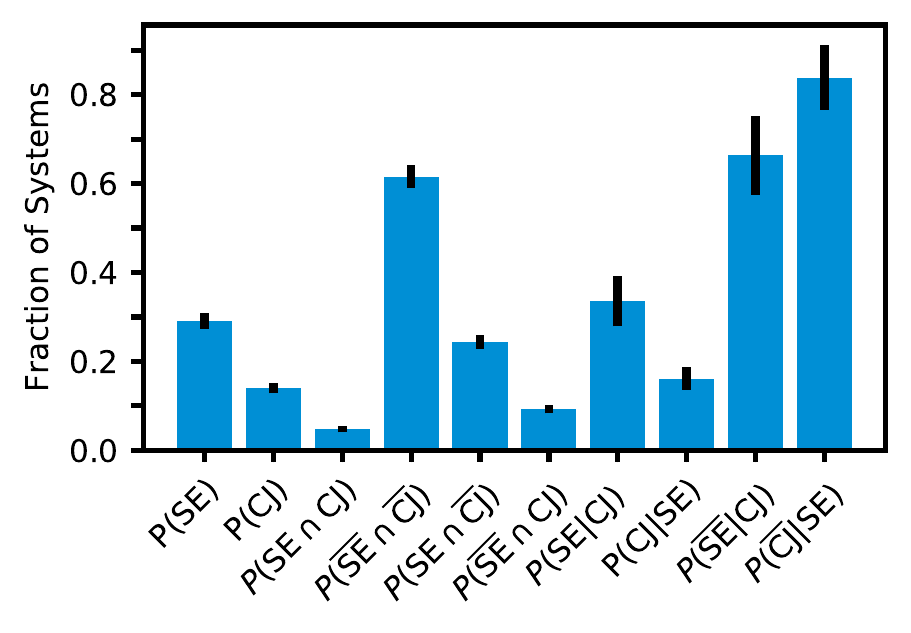}
        \caption{Fractions of planet hosts in the synthetic population \rev{of survived planets}. The height of the bars represent the probabilities of column ``Full Population'' in Table~\ref{tab:probs}. Black markers denote uncertainties assuming a Poisson statistic.}
        \label{fig:probabilityBars}
\end{figure}

In Fig.~\ref{fig:probabilityBars}, we show unconditional and conditional probabilities for the existence of super-Earths and cold Jupiters. Table~\ref{tab:probs} lists their numerical values and compares them with their counterparts inferred from observations. These are based on confirmed planets that were initially detected with the RV method~\citep{Zhu2018}. Where values are missing in their paper, we compute them using standard rules of probability theory (compare Sect.~\ref{sec:occurrenceProb}). To maintain sufficient orbital separation between the considered planet classes and for easier comparison, we do not include ``Warm Jupiters'' but adhere to the criteria in Table~\ref{tab:planetTypes}.

\citet{Zhu2018b} report a super-Earth frequency of $0.30$ based on detections of the \textit{Kepler} survey. On the other hand, cold Jupiters are found in \SI{11}{\percent} of systems around solar-type hosts~\citep{Cumming2008}. \citet{Wittenmyer2016} derive a frequency of $6.2\substack{+2.8 \\ -1.6}~\SI{}{\percent}$ but consider only giants with orbital periods $\sim$\SIrange{5}{19}{yr}. %
\citet{Herman2019} find an occurrence rate of $15\substack{+6 \\ -5}~\SI{}{\percent}$ for large planets within a period range of \SIrange{2}{10}{yr} based on newly detected transit candidates from Kepler.

\begin{figure}
        \centering
        \includegraphics[width=\hsize]{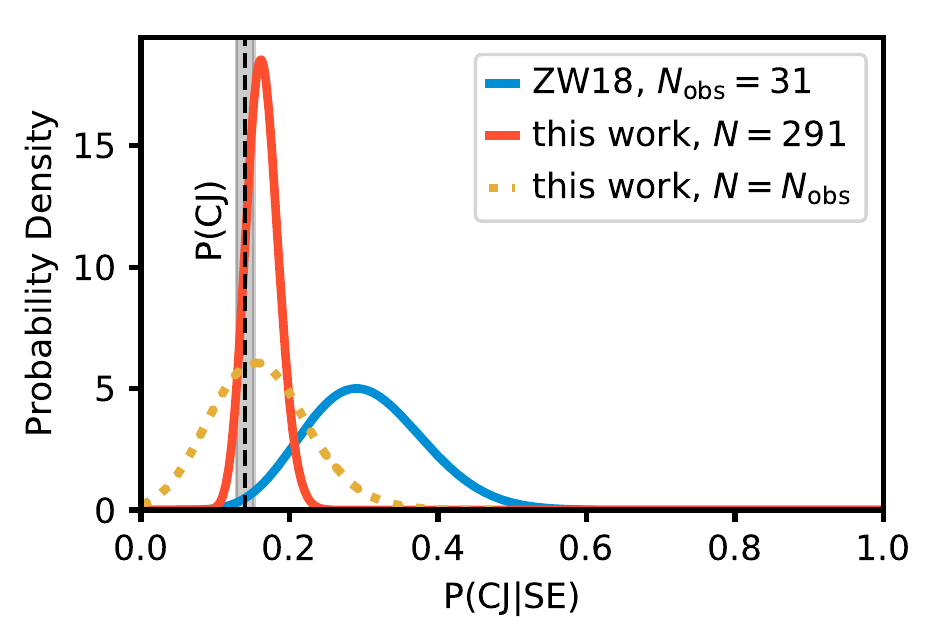}
        \caption{Observed and theoretical conditional probability P(CJ|SE). The blue curve approximates the \rev{posterior probability density to find} 9 cold Jupiter systems in a sample of $N_\mathrm{obs}=31$ super-Earth systems~\citep{Zhu2018}. Our theoretical population contains 291 systems with super-Earths, 47 of which contain cold Jupiters. The corresponding conditional probability (red curve) is enhanced compared to the overall \rev{cold Jupiter} occurrence (black dashed line).
        We further generate a KDE of P(CJ|SE) from 1000 random draws of $N_\mathrm{obs}$ synthetic super-Earth systems (dotted line). While we find lower probabilities than \citet{Zhu2018}, an anti-correlation P(CJ|SE)~<~P(CJ) is unlikely.}
        \label{fig:probDistCJSE}
\end{figure}

With $\mathrm{P(SE)_{syn}} = 0.29\pm0.02$ and $\mathrm{P(CJ)_{syn}} = 0.14\pm0.01$, our synthetic population is consistent with these observables, although it contains slightly more cold Jupiters. Solar System analogs, that is, systems containing a cold Jupiter but lacking super-Earths, are rare ($\mathrm{P(\overline{SE} \cap CJ)_{syn}} = \SI{0.09\pm0.01}{}$). This quantity is difficult to constrain observationally, but \citet{Zhu2018} give an order-of-magnitude estimate of \SI{1}{\percent}.
More interesting for the relation between inner rocky planets and outer giants are the conditional probability of having at least one cold Jupiter in a super-Earth-hosting system, P(CJ|SE), and its inverse P(SE|CJ). \citet{Zhu2018} found nine cold Jupiter-hosting systems in $N_\mathrm{obs}=31$ super-Earth systems\footnote{for consistency, we consider only their sample that excludes warm Jupiters} and thus report \rev{$\mathrm{P(CJ|SE)_{obs}} = 9/31 \approx 0.29$}. The result is supported by \citet{Bryan2019}, who find that \SI{39\pm7}{\percent} of systems with inner super-Earths (\SIrange{1}{4}{R_\oplus}, \SIrange{1}{10}{M_\oplus}) host an outer gas giant. Both studies conclude that, compared to field stars, cold Jupiters are more prevalent around stars hosting super-Earths at short orbital distances. In our synthetic population, 47 out of 291 super-Earth systems contain at least one cold giant, which yields a rate of $\mathrm{P(CJ|SE)} = 0.16\pm0.03$.
\rev{In Fig.~\ref{fig:probDistCJSE}, we construct the binomial likelihood of this result.
The distributions shown describe the probabilities to find $k$ CJ systems when we randomly draw $N$ SE systems.
We compare the rate from our simulations (red curve, $k=47, N=291$) to the ones found in \citet{Zhu2018} (blue curve, $k_\mathrm{obs}=9, N_\mathrm{obs}=31$).}
While the distributions have significant overlap, our result lies $1.7$ standard deviations from the observed one. Despite these differences, the anti-correlation case P(CJ|SE) $<$ P(CJ) remains unlikely.

We further estimated the variance in $\mathrm{P(CJ|SE)}$ we would expect if our sample size of super-Earth systems was the same as the observed one, $N=N_\mathrm{obs}$. The dotted line in Fig.~\ref{fig:probDistCJSE} shows the corresponding kernel density estimation (KDE) from 1000 random draws. Its standard deviation is $0.06$ and the probability to find exactly nine cold Jupiter systems is 3~\SI{}{\percent}.

\rev{The inverse conditional probability of finding a super-Earth in a cold Jupiter hosting system, P(SE|CJ), could observationally only be constrained using indirect methods.
\citet{Zhu2018} derived it using Bayes' law and assumptions on the individual detection probabilities P(SE) and P(CJ).
They found $\mathrm{P(SE|CJ)_{obs}} = 90\pm20\,\%$.
In the synthetic case, we can measure this quantity directly and obtain a much lower probability of $0.34\pm 0.06$.
This result differs from the non-correlation case P(SE|CJ) $=$ P(SE) by $1.1$ standard deviations, suggesting that the occurrence of super-Earths is slightly enhanced in cold Jupiter hosting systems compared to field stars.}

\subsection{Removal of super-Earths}\label{sec:SEremoval}
We find 93 systems that contain cold Jupiters but no super-Earths after \SI{5e9}{yr}. This raises the question if the latter
a) never existed; or b) disappeared during the formation phase.

All simulations start with 100 planet seeds of \SI{0.01}{M_\oplus} each. Therefore, if the first hypothesis is true, we expect a significantly increased number of planets that had their growth stalled before reaching super-Earth mass, that is, due to the competition between solid material with other planets (particularly giants).

\rev{Thus, we confronted the fractions of planets in cold Jupiter-hosting systems without super-Earths ($\mathrm{\overline{SE} \cap CJ}$) and with super-Earths ($\mathrm{SE \cap CJ}$), respectively.
To facilitate a comparison with the observed sample, we only used planets with $M_\mathrm{P} \sin i \leq \SI{2}{M_\oplus}$.
}
In order to avoid biases introduced by planets that were accreted \rev{by the star} or ejected \rev{from the system}, we counted all planets in this mass range regardless of their ultimate fate. We find that the fraction of failed super-Earths is %
the same in both populations (0.80 compared to 0.81)\footnote{We caution that most of these planets retained masses close to our initial embryo mass; therefore these values should not be mistaken as occurrence rates for low-mass planets.}. %
This shows that planetary growth to (at least) super-Earth mass was not inhibited in $\mathrm{\overline{SE} \cap CJ}$ systems and hypothesis \led{a)} must be rejected.

To address hypothesis \led{b)}, we distinguish between three scenarios that remove planets after they formed. They can be ejected out of the system, become accreted by the host star, or merge with another planet. Of the removed super-Earths in $\mathrm{\overline{SE} \cap CJ}$ systems, 29~\SI{}{\percent} are ejected, 11~\SI{}{\percent} are accreted by the star, and 60~\SI{}{\percent} are accreted by another planet.

Comparing the frequency of ejections in different populations, we find that from almost all (\SI{99}{\percent}) $\mathrm{\overline{SE} \cap CJ}$ systems a planet in the super-Earth mass range (compare Table~\ref{tab:planetTypes}) was ejected.
This compares to a significantly smaller fraction of 19~\SI{}{\percent} for the overall population. Furthermore, \SI{22}{\percent} of all super-Earths in systems hosting cold Jupiters were ejected, while only \SI{2}{\percent} of super-Earths in non-cold Jupiter systems were ejected.
The fraction of super-Earths that become accreted to the host star is small and comparable across the different populations, regardless of the presence of giants in the system.

An equally catastrophic \led{and} more common destiny for growing super-Earths are collisions with other protoplanets. During such events, part or all of the mass of a planet is transferred to the collision partner. In the majority of cases, this partner is a roughly terrestrial-mass body; only 20~\SI{}{\percent} of events in the $\mathrm{\overline{SE} \cap CJ}$ population correspond to a giant-mass partner. However, the ``winner'' of such a collision is likely to experience another planetary encounter during its lifetime, possibly with destructive consequences. We traced each accreted planet through its entire subsequent collisional history to determine which planet in the system became the final recipient of its material. In the $\mathrm{\overline{SE} \cap CJ}$ population, only \SI{26}{\percent} of these final accretors were one of the cold Jupiters in the system. Eventually, \SI{34}{\percent} of accreted planets end up in super-Earth-mass planets (that might not survive), and \SI{31}{\percent} in planets less massive than \SI{2}{M_\oplus}.
We conclude that hypothesis~\led{b)} is correct and the majority of missing super-Earths merged with another planet in their system.

\begin{figure*}
        \centering
        \includegraphics[width=\hsize]{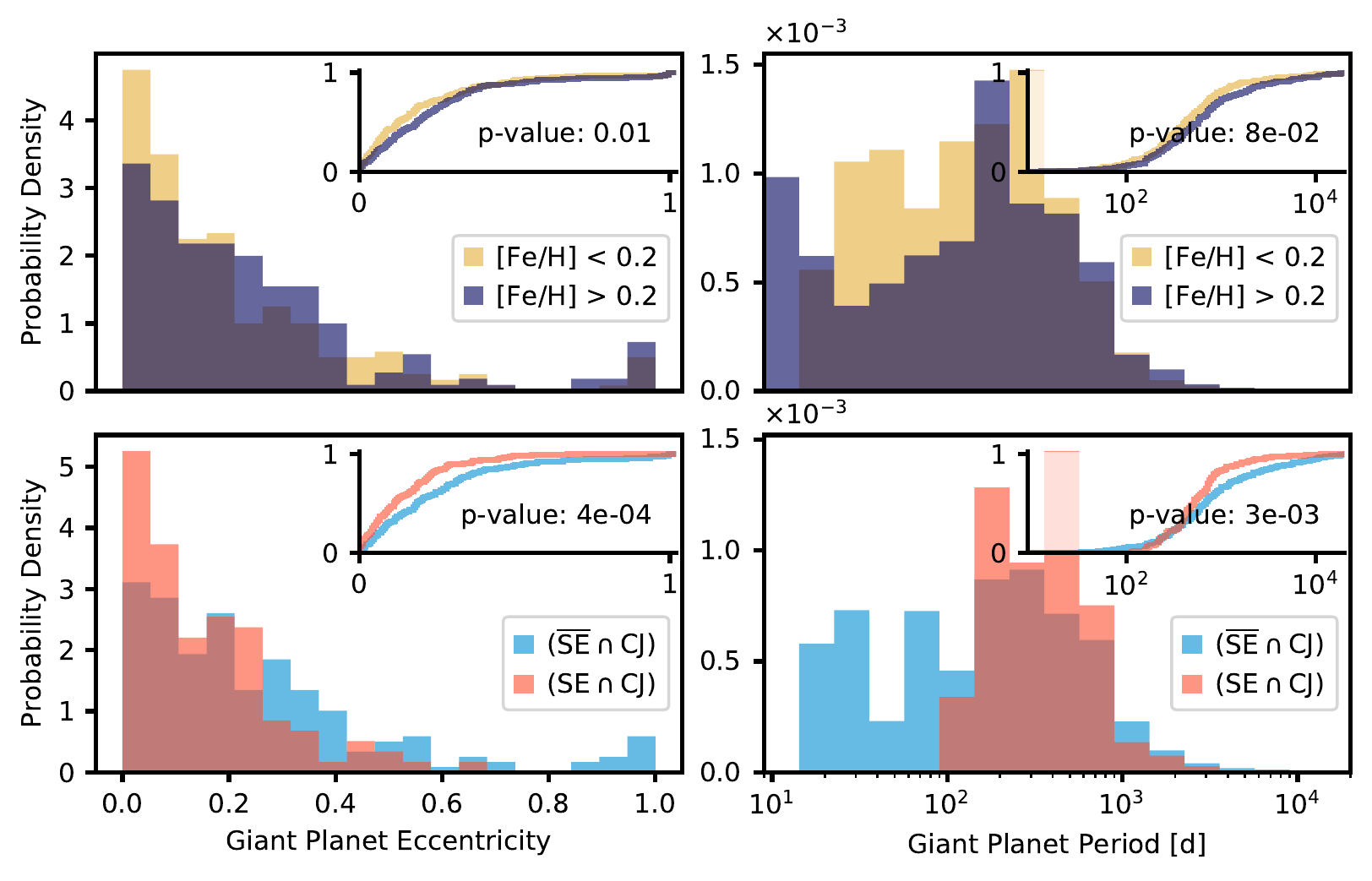}
        \caption{Eccentricity and period distributions of all giant planets that ever formed, regardless of their survival. This plot includes all planets with masses from \SI{95}{M_\oplus}, not only cold Jupiters. %
Insets show the corresponding empirical distribution functions.\protect\\
        \textit{Upper left}: Eccentricity distribution for giants orbiting low-metallicity ([Fe/H] < 0.2, yellow) and high-metallicity ([Fe/H] > 0.2, blue) host stars. Planets with very low eccentricity are slightly more prevalent in low-metallicity systems.\protect\\ %
        \textit{Upper right}: Period distributions of giant planets for different metallicities.
        A population of giants with very short periods exists only in the high-metallicity sample ($p=\SI{8e-2}{}$).\protect\\
        \textit{Lower left}: Eccentricity distributions of giant planets in cold Jupiter-hosting systems with and without super-Earth companions. Giants in systems without super-Earths have significantly higher eccentricities ($p=\SI{5e-4}{}$). No super-Earths occur when a giant with $e \gtrsim 0.7$ exists.\protect\\
        \textit{Lower right}: Period distributions of giants with and without super-Earths. The latter persist only in systems without short-period giants.}
        \label{fig:CJdistributions}
\end{figure*}

\begin{figure}
        \centering
        \includegraphics[width=\hsize]{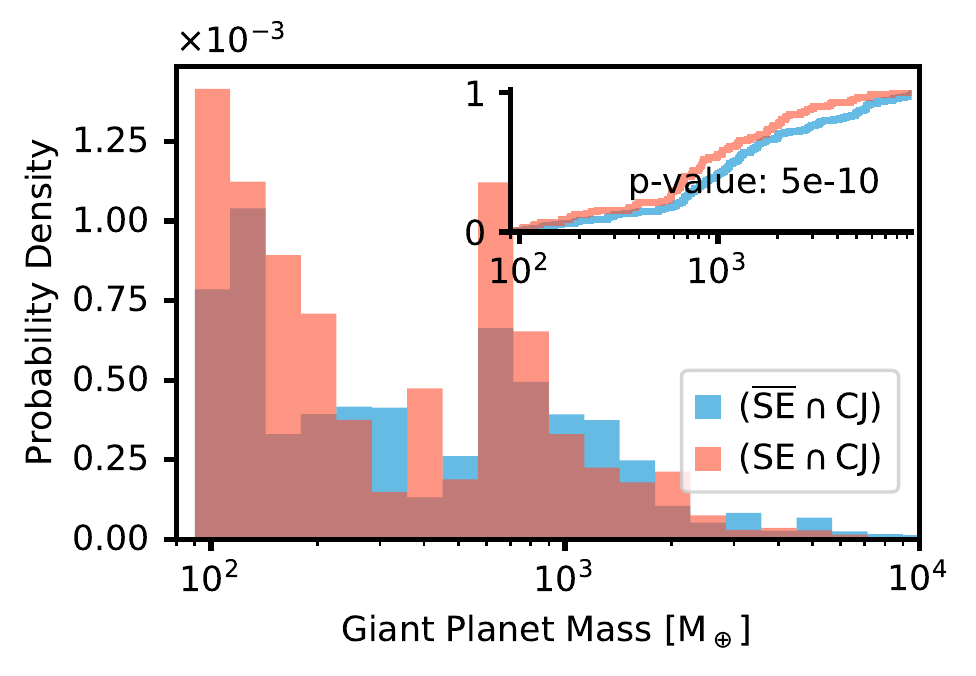}
        \caption{Mass distribution of giant planets in cold Jupiter-hosting systems with and without super-Earths. The mass distributions of the two populations show only minor differences.
        }
        \label{fig:giantMasses_histCDF}
\end{figure}

In the following, we investigate the cause of these merger events.
In the lower panel of Fig.~\ref{fig:CJdistributions}, we show the eccentricity and period distributions of giant planets ($M_\mathrm{P} \geq \SI{95}{M_\oplus}$) in $\mathrm{\overline{SE} \cap CJ}$ and $\mathrm{SE \cap CJ}$ systems, respectively. Giants in systems with removed super-Earths have, on average, significantly higher eccentricities ($p=\SI{5e-4}{}$). Their periods reach down to tens of days, while no giants with $P \lesssim \SI{100}{\day}$ exist in systems with super-Earths.
The planet masses follow similar distributions for both populations (Fig.~\ref{fig:giantMasses_histCDF}), but there are differences on either extreme of the distribution: while $\mathrm{SE \cap CJ}$ systems have a higher occurrence of ``Saturns'' ($M_\mathrm{P} \sim \SI{100}{M_\oplus}$), very-high-mass planets that reach into the Deuterium-burning regime occur only in the $\mathrm{\overline{SE} \cap CJ}$ population.

\begin{figure*}
        \centering
        \includegraphics[width=\hsize]{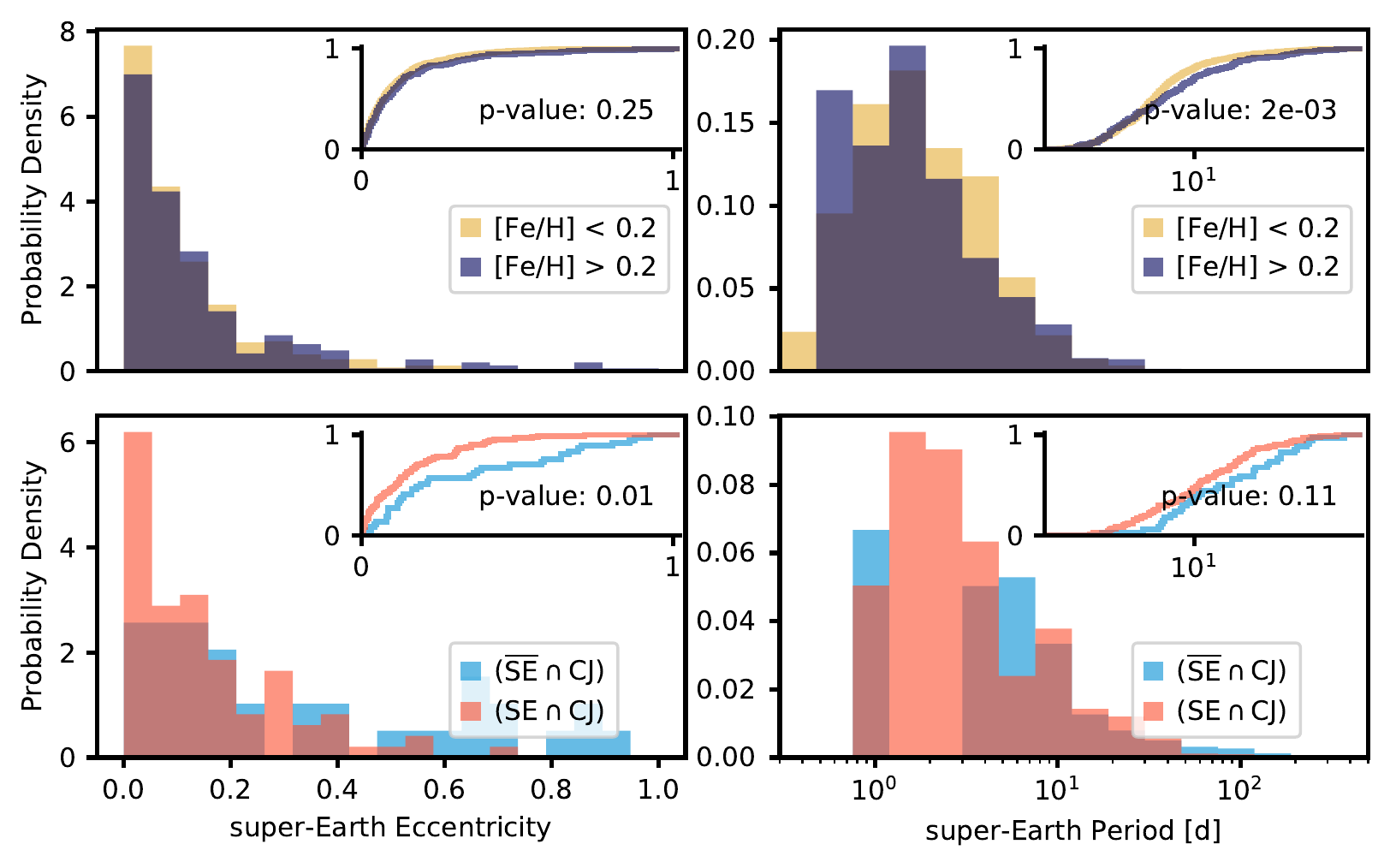}
        \caption{Same as Fig.~\ref{fig:CJdistributions}, but for all super-Earths that ever formed during the systems' histories. This includes both surviving planets and planets that we classified as super-Earths at the time of their removal (via ejection or collision events). Planets accreted to the star are not shown. %
        With $p=0.25$, a Kolmogorov-Smirnov test suggests a negligible statistical distance between the eccentricity distributions in high- and low-metallicity systems. On the other hand, the difference between systems with and without super-Earths is significant: where they are missing, eccentricities are strongly enhanced.
        The period distribution in the $\mathrm{SE \cap CJ}$ population, which contains surviving super-Earths, is shifted towards lower values compared to $\mathrm{\overline{SE} \cap CJ}$. These trends point to dynamical excitation of super-Earths by giant companions where they are present.
        }
        \label{fig:quadFig_SEeverLived}
\end{figure*}

Figure~\ref{fig:quadFig_SEeverLived} shows the same distributions for planets in the super-Earth mass range. We included not only planets that survive the entire formation and evolution phase, but all planets that reached SE mass and never grew beyond.
Planets that were accreted onto the stars where removed since the physical meaning of their final period and eccentricity is questionable. On average, $\mathrm{\overline{SE} \cap CJ}$ planets have higher eccentricities and larger periods than $\mathrm{SE \cap CJ}$ planets. Ultra-short periods of less than three days are rare in $\mathrm{\overline{SE} \cap CJ}$ systems. %

\subsection{Ice mass fractions}\label{sec:iceMassFractions}

While detailed analyses of planetary compositions are beyond the scope of this paper, we modeled the abundances of relevant chemical species and took into account the condensation of volatiles as a function of radial distance~\citep{Thiabaud2015}.

\begin{figure}
        \centering
        \includegraphics[width=\hsize]{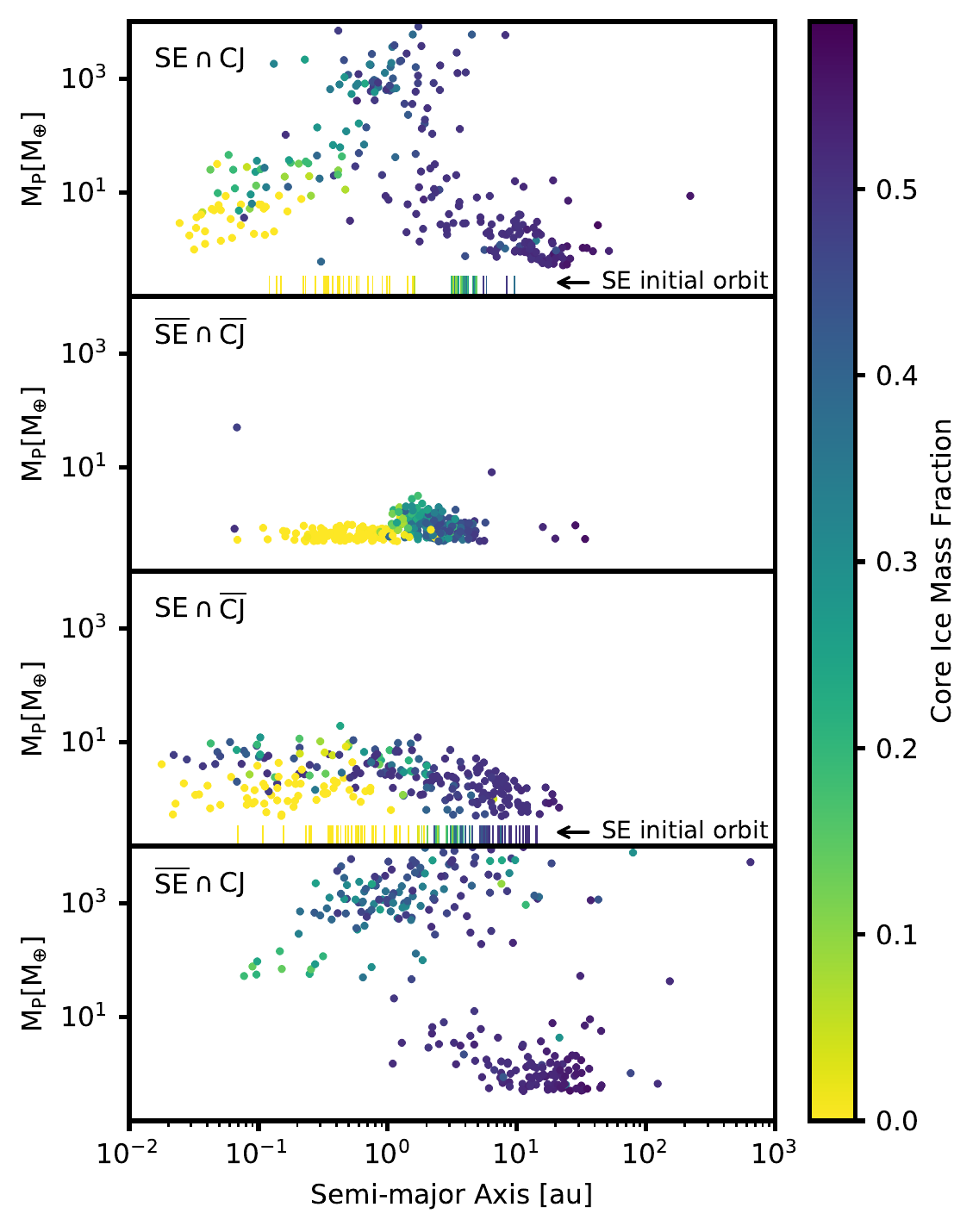}
        \caption{Water ice mass fractions of planet cores in the different system classes. For balanced samples of 264 planets per class, we show their position in mass-semi-major axis space, color-coded by ice mass fraction.
        For the systems containing super-Earths, we indicate their initial orbital distance by a rug plot with the same color-code. Planets $< \SI{0.5}{M_\oplus}$ are not shown.
        The mass fractions of ice in the core are largely determined by the position of the water ice line in the protoplanetary disk, where planets beyond $\sim \SI{1}{au}$ are mostly water-rich. Super-Earths in systems without cold Jupiters have higher ice mass fractions than their siblings with giants.
        }
        \label{fig:iceFractionScatter}
\end{figure}

To avoid distortions introduced by inclination effects and detection bias, we adopted for the planet classification into super-Earths and cold Jupiters an approach that reflects the architectures of our synthetic systems better than the limits in Table~\ref{tab:planetTypes}.
\rev{This includes different mass limits for the inner planets, which are relatively abundant at higher masses than the \SI{20}{M_\oplus} limit we used for the comparisons above (compare Fig.~\ref{fig:occMapMasses}).
We chose an upper limit of \SI{47}{M_\oplus} (half of the lower limit for giant planets) for these planets.
}
For each system, we:
\begin{enumerate}
        \item checked if a giant planet exists, using our nominal mass limits. If yes, the upper period limit for super-Earths equals the period of the innermost giant. Otherwise, we used the same limit as in Table~\ref{tab:planetTypes} of $P = \SI{400}{\day}$.
        \item set mass limits for \led{``massive''} super-Earths of $\SI{1}{M_\oplus} < \mathrm{M_P} < \SI{47}{M_\oplus}$
        \item did not impose a detection limit
        \item did not draw an inclination term $\sin i$ but used the planets' mass $\mathrm{M_P.}$
\end{enumerate}
Using these rules, we classified the population of systems into four distinct classes $\mathrm{(SE \cap CJ)}, \,\mathrm{(\overline{SE} \cap \overline{CJ})}, \,\mathrm{(SE \cap \overline{CJ})}$, and $\mathrm{(\overline{SE} \cap CJ)}$.
For each of these system classes, we tracked the water content of individual planet cores throughout their formation history and show their ice mass fractions at $t = \SI{5}{\giga yr}$ in Fig.~\ref{fig:iceFractionScatter}.

The individual panels show balanced samples of 264 planets more massive than \SI{0.5}{M_\oplus} from systems containing different combinations of super-Earths and cold Jupiters.
The sample size corresponds to the number of planets in the smallest class.

\begin{figure}
        \centering
        \includegraphics[width=\hsize]{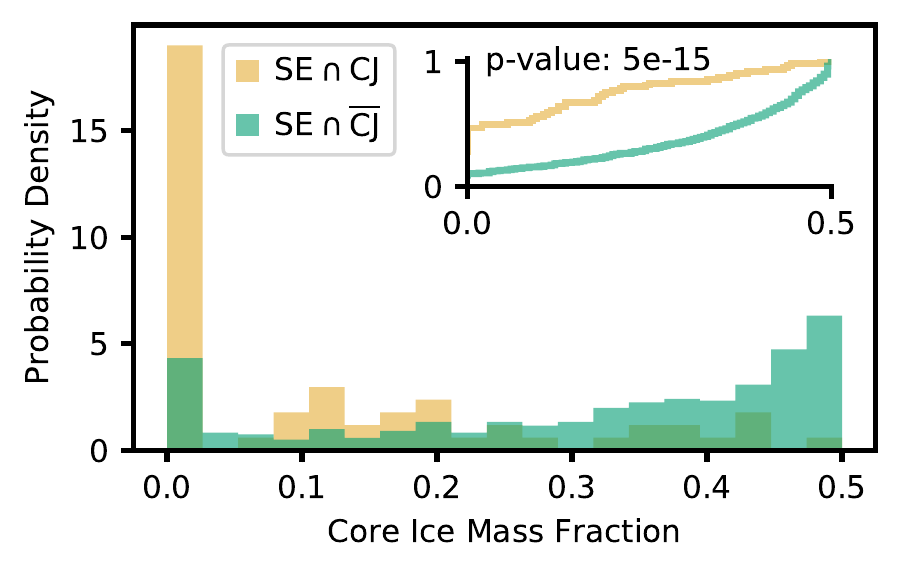}
        \caption{Distribution of ice mass fractions in the cores of super-Earths with and without cold Jupiter companion.
    The vast majority of super-Earths with giant companions is completely dry.
        On the other hand, super-Earths in systems without a giant planet often retain large ice mass fractions, reaching close to the maximum value of $f_\mathrm{ice}\approx 0.59$.
        With $p\approx \SI{e-15}{}$, the null hypothesis that both datasets are drawn from the same parent population must be rejected.
        }
        \label{fig:iceMassFractions_histCDF}
\end{figure}

The core ice mass fractions $f_{\mathrm{ice}}$ of growing super-Earths are mainly determined by their initial orbital distances, which are indicated by the rug plot at the bottom of each panel.
The colors of the markers correspond to the final ice mass fractions and show their strong correlations with the planet embryo positions relative to the water ice line.
Cores that start outside of the ice line are strongly enhanced in water ice.
Mixing of planetary compositions at later times is only possible due to different migration trajectories, collisions, and scattering events.
Planets beyond $\sim \SI{1}{au}$ almost always maintain their high ice mass fraction. On the other hand, more close-in planets show varying compositions.
On average, super-Earths in systems without outer giants are more ice-rich than their siblings in cold Jupiter-hosting systems.
They mostly started just outside the water ice line and then migrated to their final positions, while super-Earths in cold Jupiter-hosting systems usually start within the ice line.
The distributions of $f_{\mathrm{ice}}$ differ significantly between the two populations (compare Fig.~\ref{fig:iceMassFractions_histCDF}).
The median ice mass fractions of the two super-Earth-hosting classes amount to $f_{\mathrm{ice},\, \mathrm{SE \cap \overline{CJ}}} =  0.23 \substack{+0.27 \\ -0.23}$ and $f_{\mathrm{ice},\, \mathrm{SE \cap CJ}} =  0.02 \substack{+0.29 \\ -0.02}$, respectively, where upper and lower bounds denote the 84th and 16th percentiles.

\begin{figure}
        \centering
        \includegraphics[width=\hsize]{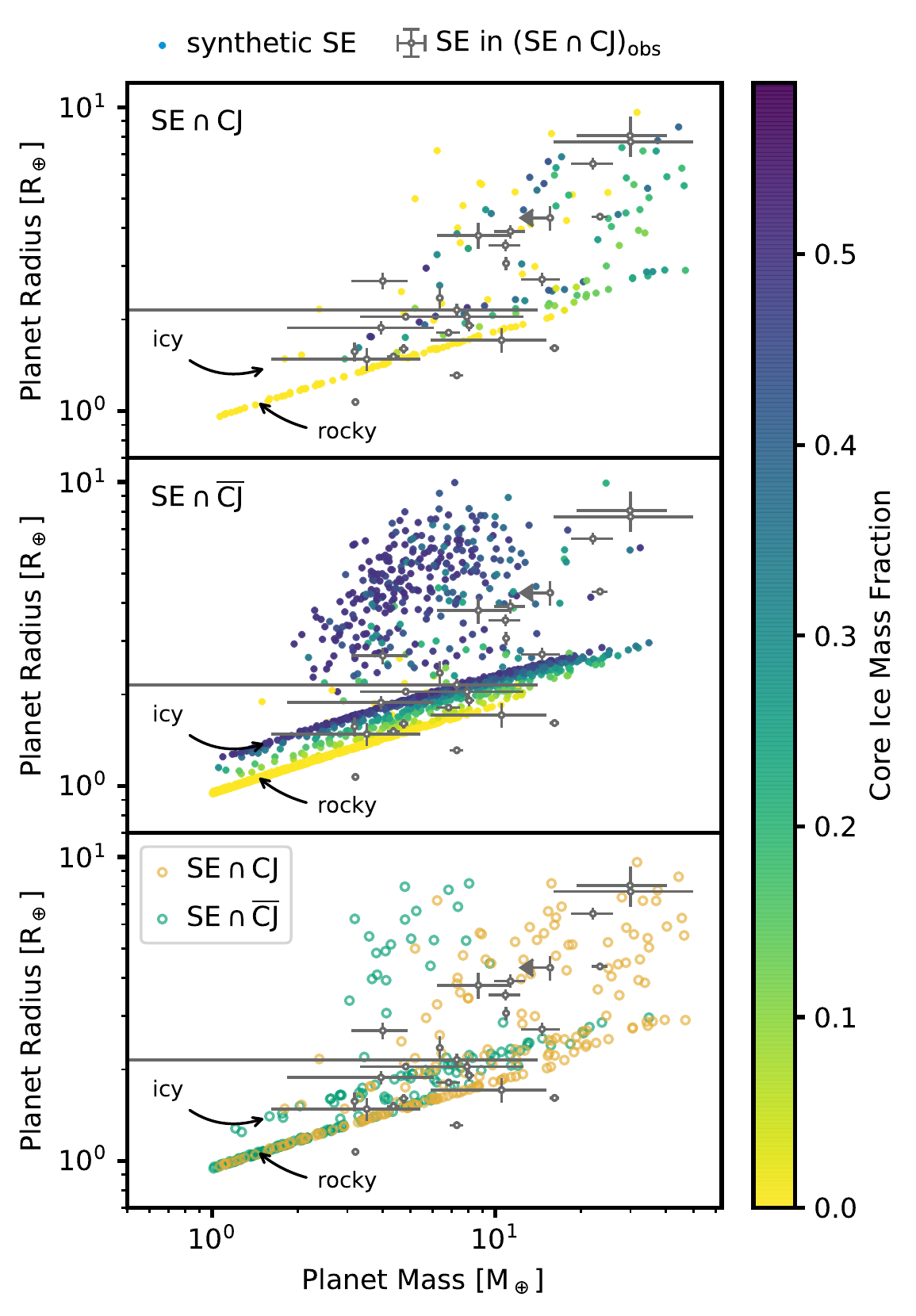}
        \caption{Mass-radius diagram of inner super-Earths with and without cold Jupiters.
        Included are all planets \rev{with masses of \SIrange{1}{47}{M_\oplus} and $a < \SI{0.3}{au}$.} The top and center panels show all such planets in  $\mathrm{SE \cap CJ}$ and $\mathrm{SE \cap \overline{CJ}}$ systems, respectively, with their core ice mass fractions color-coded.
        The bottom panel shows balanced samples (N=190) from both system classes.
        On average, super-Earths in cold Jupiter-hosting systems populate regions of higher bulk density.
        In all panels, we overplot observed super-Earths in exoplanet systems containing both planet types, $\mathrm{(SE \cap CJ)_{obs}}$.
        These planets match their synthetic counterparts well, but fail to match the $\mathrm{SE \cap \overline{CJ}}$ population.
        }
        \label{fig:mass-radius_CJ-noCJ}
\end{figure}

These differences in composition are reflected in different bulk densities, which can be probed in a mass-radius diagram (Fig.~\ref{fig:mass-radius_CJ-noCJ}).
Here, we consider all inner ($a < \SI{0.3}{au}$) planets with masses \SIrange{1}{47}{M_\oplus}.
In general, three groups of planets can be identified:
\begin{itemize}
        \item Rocky planets without significant gaseous envelopes (lower diagonal chain of markers). These planets occupy the diagram areas with the highest densities and are the most abundant group.
        \item Icy planets without significant gaseous envelopes (upper diagonal chain of markers). Planets of this group have slightly lower bulk densities due to their high ice mass fractions.
        \item Planets that accreted and maintained significant atmospheres. These envelope-dominated planets are clearly detached towards larger radii.
\end{itemize}

As is apparent in the plot, cold Jupiter-hosting systems (top panel) are almost completely free from icy, atmosphere-less super-Earths.
On the other hand, systems without cold Jupiters (center panel) are mainly populated by super-Earths with ice-rich cores.
This holds also for planets with significant H/He envelopes, which puts them to lower average masses.

In the bottom panel, $\mathrm{SE \cap CJ}$ and $\mathrm{SE \cap \overline{CJ}}$ systems are shown in different colors.
Here, the sample sizes of the different classes are balanced.
A clear difference between the two classes is visible: super-Earths in cold-Jupiter hosting systems show larger bulk densities, whereas those without a giant companion tend to populate regions of less density.
The effect is even stronger for planets with large radii, that is, significant H/He envelopes, where those with masses below $\sim \SI{10}{M_\oplus}$ exist almost only in the class without cold Jupiters.

These differences pose an interesting prediction for exoplanets with both mass and radius measurements.
For a first-order comparison with real exoplanets, we constructed a sample of observed systems containing both super-Earths and outer giants, $\mathrm{(SE \cap CJ)_{obs}}$.
We did not, however, go so far as to compose an observed counterpart for the $\mathrm{SE \cap \overline{CJ}}$ population since this sample would suffer from a strong bias: a system where no cold Jupiter was detected is not guaranteed to contain no such planets.
The observed sample was constructed as follows: we obtained from the NASA exoplanet archive\footnote{queried on 2020-03-18} all confirmed planets and
classified them in the same way we did for the synthetic population, using our flexible period limits (see above).
We then kept only systems that contain both super-Earths and cold Jupiters.
From these systems, we include the 26~super-Earths that have both mass and radius measured in the mass-radius diagram (Fig.~\ref{fig:mass-radius_CJ-noCJ}).
This confrontation with the theoretical sample reveals a remarkable agreement of $\mathrm{(SE \cap CJ)_{obs}}$ with its synthetic counterpart, especially in the regime of planets with significant atmospheres.
Here, $\mathrm{(SE \cap CJ)_{obs}}$ differs substantially from the synthetic $\mathrm{SE \cap \overline{CJ}}$ sample.
In particular, $\mathrm{(SE \cap CJ)_{obs}}$ matches the synthetic super-Earths with cold Jupiter companions $\mathrm{SE \cap CJ}$ much better than it matches the overall super-Earth sample (compare bottom panel of Fig.~\ref{fig:mass-radius_CJ-noCJ}).

\subsection{Host star metallicity: effects on planet occurrences}\label{sec:metallicityOccurrences}
Some planet formation models suggest that cold Jupiters and super-Earths form in different metallicity domains: host stars of low metallicity are not able to provide enough solid material for giant formation but can produce super-Earths~\citep{Ida2004}. On the other hand, high-metallicity hosts with larger solid reservoirs yield giants which prevent the formation of super-Earths~\citep{Izidoro2015}.   However, \citet{Zhu2019b} not only find the presence of super-Earths in observed high-metallicity systems but even a weak positive correlation of their occurrence.

To identify trends connected to host star metallicity, we compute all above probabilities not only for the full population but also for three distinct metallicity ranges:  $[\mathrm{Fe/H}] < -0.2$; $ -0.2 < [\mathrm{Fe/H}] < 0.2$; and $0.2 < [\mathrm{Fe/H}]$. The distribution of metallicities in our synthetic population correspond to the observed values in the Solar neighborhood (see Appendix~\ref{sec:hostMet}).

While observational studies revealed only a weak dependence of super-Earth occurrence on host star metallicity~\citep[e.g.,][]{Wang2015a, Zhu2016}, we see a notable absence of super-Earths around low-metallicity stars with a super-Earth fraction of only $0.13$ compared to $0.29$ which we obtain for the full population.

Figure~\ref{fig:CJdistributions} shows a histogram and empirical distribution function of planet periods for two giant planet samples of low metallicity ([Fe/H] < 0.2) and high metallicity ([Fe/H] > 0.2). Host stars of high metallicity produce a rather bimodal distribution with a low-period bump.
A similar feature was observed in the giant exoplanet population~\citep{Santos2006}.
This bimodality does not exist in the low-metallicity sample. A two-sample Kolmogorov-Smirnov test on the period distribution yields $p=\SI{8e-2}{}$, allowing to reject the null hypothesis that the two samples stem from the same distribution.

\begin{figure*}
        \centering
        \includegraphics[width=\hsize]{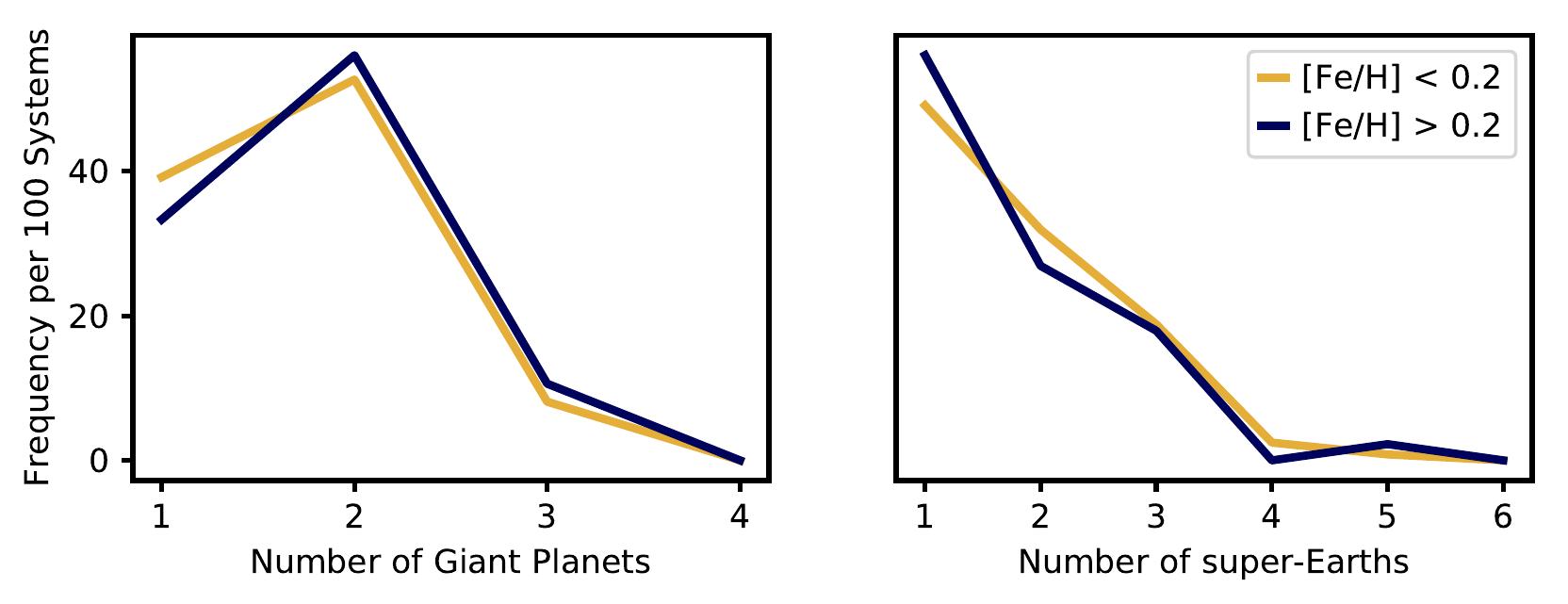}
        \caption{Multiplicity for different host star metallicities. \textit{Left:} Number of giant planets per giant-hosting system. While high-metallicity systems have marginally higher multiplicity, there are no significant differences.\
\textit{Right:} Number of super-Earths per \rev{super-Earth hosting system}. Again, the differences are statistically indistinguishable.}
        \label{fig:NgiantsSE_met}
\end{figure*}

\begin{figure}
        \centering
        \includegraphics[width=\hsize]{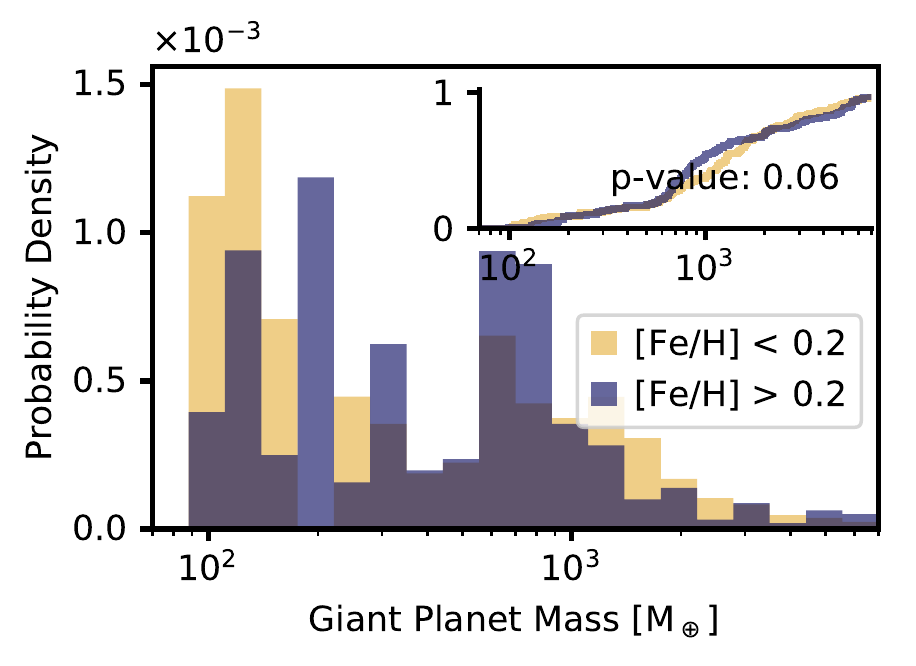}
        \caption{Mass distribution of giant planets in low- and high-metallicity systems. The difference between the distributions is statistically indistinguishable.}
        \label{fig:MassGiants_met}
\end{figure}

A comparison of the number of giant planets per system between these samples (Fig.~\ref{fig:NgiantsSE_met}) shows no significant difference: where giant planets occur, their multiplicities follow the same distribution regardless of the metallicity.
There is also no significant change in the planet mass distribution with metallicity, as has already been found in~\citet{Mordasini2012d}.
As shown in Fig.~\ref{fig:MassGiants_met}, there is a difference in the CDF around $~\SI{1000}{M_\oplus}$, but with $p=0.06,$ we cannot exclude equal source distributions.

\subsection{Possibility of reduced multiplicity in cold Jupiters}\label{sec:multiplicity}
In this study, we denote as multiplicity, $\mu,$ the number of planets in a given system.
Where we quote mean multiplicities across systems $\bar{\mu}$, we consider all planets that are above our detection limit of $K = \SI{2}{\meter\per\second}$ (compare Sect.~\ref{sec:detectionLimit}).
For the mean multiplicity of a specific planet type, we consider only systems containing at least one such planet.

\citet{Hansen2017} suggested that dynamically hot outer giants can perturb inner terrestrial planets and decrease the multiplicities of these systems. Support for this hypothesis came from \citet{Zhu2018}, who find a deficiency of high-multiplicity systems in their cold Jupiter-hosting population, albeit with little significance due to the small sample.

\begin{figure}
        \centering
        \includegraphics[width=\hsize]{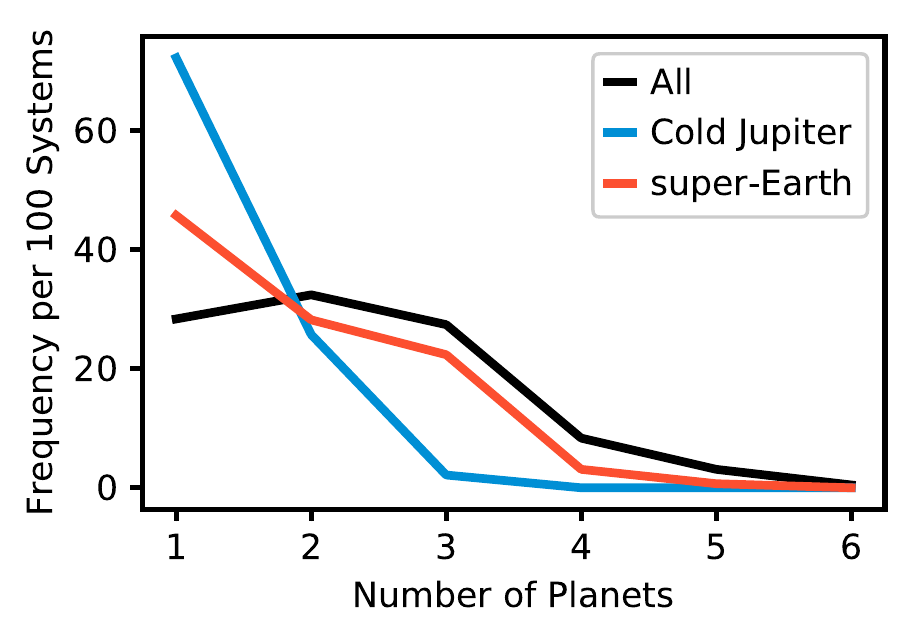}
        \caption{Normalized frequency of multiplicities for different planet types. The term \led{``All''} (black) includes all planets above our detection limit. For each of these multiplicities, we count the number of systems with this multiplicity and normalize it to 100 systems that host this species.
        }
        \label{fig:multiplicities}
\end{figure}

In our synthetic sample, we find mean multiplicities of $\bar{\mu} = 2.27 \pm 1.08$ for the complete planet sample, $\bar{\mu}_\mathrm{SE} = 1.85 \pm 0.92$ for super-Earths in super-Earth hosting systems, and $\bar{\mu}_\mathrm{CJ} = 1.30 \pm 0.50$ for cold Jupiters in cold Jupiter hosting systems, quoting arithmetic mean and standard deviation.
The frequency of multiplicities for different planet types is depicted in Fig.~\ref{fig:multiplicities}.
We note that the plot shows the frequency per 100 systems containing
the species  (i.e. the sum of frequencies for each species equals to 100), it does not, therefore, reflect overall planet occurrences.
In black, we show multiplicity frequencies for the complete sample, with orange and blue corresponding to the systems containing super-Earths and cold Jupiters, respectively.
On average, the multiplicity of super-Earths is higher than for cold Jupiters.
The latter show a multiplicity rate (fraction of systems with $\mu > 1$) of \SI{28}{\percent}, consistent with the rate for observed cold Jupiters~\citep[e.g.,][]{Wright2009}. Less than \SI{4}{\percent} of all systems show a multiplicity greater than four and no systems with $\mu > 5$ exist in the population.

\begin{figure}
        \centering
        \includegraphics[width=\hsize]{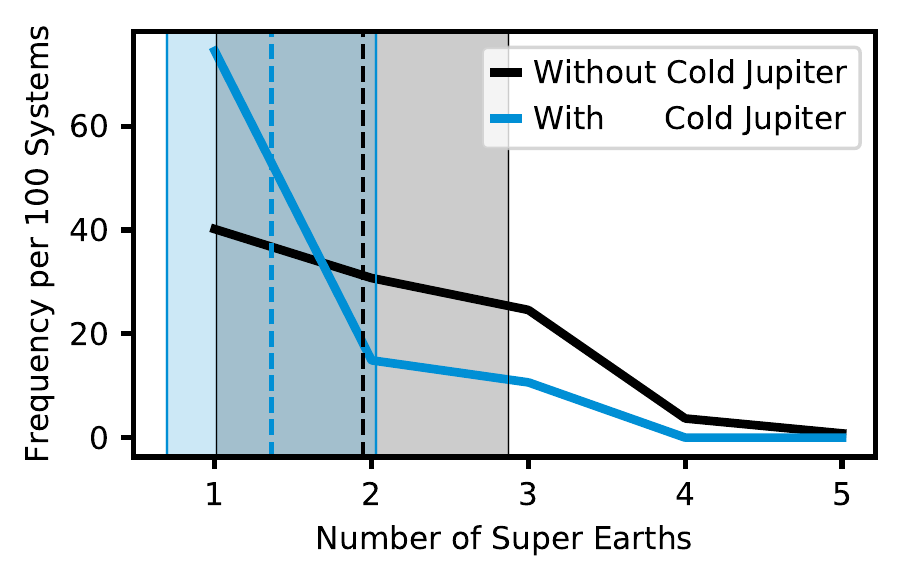}
        \caption{Normalized frequency of super-Earth multiplicity in systems with and without cold Jupiter. In accordance with Fig.~\ref{fig:multiplicities}, the frequency is normalized to 100 systems of the respective subpopulation. Dashed lines show mean multiplicities, whereas shaded regions reach to one standard deviation from the mean. super-Earth multiplicity is \rev{enhanced} in systems without outer companions.
        }
        \label{fig:multiplicitiesCJnoCJ}
\end{figure}
To investigate a possible influence of the presence of an outer giant on the multiplicity of inner terrestrial planets, we examine the subpopulations of cold Jupiter systems and non-cold Jupiter systems separately. Figure~\ref{fig:multiplicitiesCJnoCJ} compares the frequency of super-Earth multiplicities for these two samples. Again, frequencies are normalized to 100 systems of the respective subpopulation. The mean super-Earth multiplicity is \rev{slightly} enhanced in systems without outer companions ($1.94 \pm 0.93$) compared to systems hosting a cold Jupiter ($1.34 \pm 0.67$). %

\rev{This is consistent with the observed trend of reduced multiplicity, but the effect is not significant enough to confirm it.}
\rev{We do not observe a } difference in super-Earth multiplicities of low-metallicity systems ([Fe/H] < 0.2) and high-metallicity systems ([Fe/H] > 0.2) (see Fig.~\ref{fig:NgiantsSE_met}), as was suggested by \citet{Brewer2018}.

\section{Discussion}\label{sec:discussion}

\subsection{\rev{Observed and theoretical host star fractions}}

While the absolute fractions of super-Earth hosts and cold Jupiter hosts in our synthetic population are largely consistent with observations, there are considerable differences in the conditional probabilities $\mathrm{P(CJ|SE)}$ and $\mathrm{P(SE|CJ)}$.
Observationally, only the former can be directly derived from counting statistics without additional assumptions on $\mathrm{P(CJ)}$ and $\mathrm{P(SE)}$.

\citet{Zhu2018} reported 9 out of 31 super-Earth systems that host a cold Jupiter, which results in a range of probabilities of $\mathrm{P(CJ|SE)_{obs}} = 0.29\pm0.18$ (compare Fig.~\ref{fig:probDistCJSE}). This is in line with \citet{Bryan2019}, who found that super-Earth systems are enhanced in probability of also hosting a long period giant based on a sample of 65 stars.
\citet{Zhu2018} based their study on RV detections and \citet{Bryan2019} \rev{used} a mixed sample of RV and transit-detected systems.
Similarly, \citet{Herman2019} analyzed transiting planets and found a positive correlation between short- and long-period planets in a sample of 12 candidate systems. They conclude that outer giants occur exclusively in systems containing smaller inner planets.

With 1000 systems, our synthetic sample is significantly larger and can thus constrain the probability density tighter than the observations to $\mathrm{P(CJ|SE)_{syn}} = 0.16\pm0.03$.
This puts our result between the non-correlation case and the strong positive correlations presented in \citet{Zhu2018} and \citet{Bryan2019}.
Our figure is in agreement with the finding of \citet{Zhu2018} within $1.7 \sigma$.

In contrast to the quantities above, the observed probability of finding a super-Earth in a system that hosts cold Jupiters, P(SE|CJ), could not be measured directly in these studies due to the poorly constrained detection bias for super-Earths.
\citet{Zhu2018} derived it from P(SE), P(CJ), and P(CJ|SE) using Bayes' theorem and report $\mathrm{P(SE|CJ)_{obs}} = 0.90\pm0.20$. \citet{Bryan2019} follow the same approach and, while not precisely constraining P(SE|CJ), come to the same conclusion.
Both results suggest that nearly all cold Jupiters are accompanied by inner super-Earths.
It is thus surprising that \citet{Barbato2018}, who conducted a survey to search for super-Earths, found zero planets in their sample of 20 cold Jupiter-hosting systems.
\rev{In \citet{Barbato2020}, it was clarified that the estimated sensitivity of the survey was inaccurate, which provides the possibility of a higher occurrence rate for inner super-Earths than the one reported.}
See Appendix~\ref{sec:Barbato} for an assessment \rev{of their null result.}

The corresponding quantity in our synthetic population is significantly lower with  $\mathrm{P(SE|CJ)_{syn}} = 0.26\pm0.05$, indicating a lack of super-Earths in our cold Jupiter-hosting systems.
In general, we do not find the reported strong positive correlation between the two planet types in our synthetic population. In the following, we discuss possible reasons for this apparent disagreement.

One explanation for the discrepancy is selection bias.
We realized that the mass-period cuts chosen by \citet{Zhu2018}, which we adopted for our comparison, are not well reflected in the architectures of our systems but, rather, they constitute arbitrary borders in our population (compare with Fig.~\ref{fig:occMapMasses}).
In particular, our population of giant planets occupies both sides of the period limit and is located on closer orbits than the observed giant exoplanet population~\citep[e.g.,][]{Fernandes2019}.
The process that is responsible for their final orbit distance is the migration efficiency, which might be overestimated in our model.
To investigate the relations between inner rocky planets and outer giants in a way that reflects our synthetic population better, we repeated the occurrence analysis with alternative mass/period limits: in each system, we first check if a giant planet exists according to the mass limits in Table~\ref{tab:planetTypes}.
If it does, we set the outer period limit for our inner super-Earths to the period of the giant closest to the star, otherwise we chose a maximum period of \SI{400}{\day}.
The planets fulfilling these period criteria are considered super-Earths if their masses obey $\SI{1}{M_\oplus} < \mathrm{M_P} < \SI{47}{M_\oplus}$. We consider the planet's actual masses instead of $\mathrm{M_P} \sin i$ and no detection limit is imposed.
Using these flexible limits, both P(CJ|SE) and P(SE|CJ) show a clear deficit compared to the respective unconditional probabilities.
Super-Earths and cold Jupiters are anti-correlated in this case.
This demonstrates that the correlations between the host star fractions are quite sensitive to the planet classification limits, which casts some doubt on the robustness of the observed trends.

At the same time, biases in the observations can falsify host star fractions, too.
Exoplanet searches in general and RV surveys in particular suffer from human interventions that distort the inferences made on the underlying exoplanet demographics:
to increase the significance of a candidate signal and to rule out false positive scenarios, it is very common to perform additional observations of a target star once such a signal emerges.
Alarmingly, this habit increases the probability of finding another planet in the same system.
The ``human intervention bias'' thus contributes to a positive correlation in the occurrence of any two planet types and in a hardly quantifiable manner, in effect.

On a similar note, the small number of considered systems in certain observational studies \citep{Zhu2018, Herman2019, Bryan2019} raises the question of whether their samples are representative of the field exoplanet population.
Undoubtedly, our synthetic population is not a perfectly accurate representation of the planetary systems in nature.
Hence, we do not claim that the observed trend stems merely from an unfavorable combination of selection and detection biases.
However, we are concerned about the sensitivity of our results on the chosen limits.
This ambiguity demonstrates yet again the importance of a thorough understanding of a sample's selection function and of its underlying biases.

\rev{Overall, } the anti-correlation in the synthetic planet population shows that giant planets on intermediate orbits can dynamically excite and ultimately destroy inner super-Earth systems \rev{\citep[also see a discussion of this scenario in][]{Masuda2019}}.
\rev{We explore this mechanism in Sect.~\ref{sec:missingSE} in more detail.}
On the other hand, if the proposed positive correlation between inner super-Earths and outer giants exists, these results \rev{might} indicate that inward migration of giant planets is not as efficient as hitherto assumed.
More sophisticated migration models that take into account multiple interacting planets are currently not available~\citep[see, however, ][]{Masset2001}.
Until they are, population syntheses with reduced migration efficiencies can possibly reconcile the observed and synthetic results.
Such simulations will test if indeed an overestimation of planet migration torques is responsible for the competition between these planet types.

\subsection{\rev{The missing super-Earths}}\label{sec:missingSE}
Compared to pebble accretion models, where the formation of inner super-Earths relies on the radial drift of roughly centimeter-sized particles~\citep{Ormel2010} that can be interrupted by an emerging outer giant~\citep{Morbidelli2015, Ormel2017, Bitsch2019a}, the individual planet cores of a system are more independent in our model which considers only accretion of planetesimals and gas.
We therefore do not expect a negative impact by massive, outer planets on the formation efficiency of cores on closer orbits.

On the other hand, such outer giant planets still block inward migration of smaller planets that formed beyond the giant's orbit.
\rev{Unlike} in the model by \citet{Izidoro2015}, super-Earths can form independently interior to any cold Jupiter companion in our simulations.
The reason is the relatively steep radial slope of the planetesimal surface density $\beta_\mathrm{s} = 1.5$.
The resulting plethora of solid material in the inner disk enables the formation of super-Earth-sized cores interior to the water ice line in disks massive enough to grow a giant planet.

Still, the observed occurrences in \citet{Zhu2018} can overall be better matched with our final population if it contained more super-Earths.
In Sect.~\ref{sec:SEremoval} we showed that these planets are not lacking because they did not form, but because they were removed at some point in the formation and evolution phase.
With smaller fractions of these failed super-Earths lost to ejections and accretion into the star, \SI{60}{\percent} disappeared in a merger event with another planet.

We consider whether the fate of the missing super-Earths was, in fact, sealed by a population of dynamically hot giants. In turning to the eccentricity distribution of giant planets (Fig.~\ref{fig:CJdistributions}), we note that the lowest eccentricities are more prevalent for planets with a super-Earth companion in the system. There is also a population of highly eccentric giants that is missing in systems with super-Earths. %
The imprint of these giants can be seen in the eccentricity and period distribution of intermediate-mass planets, which differ significantly between systems with and without super-Earths (compare Fig.~\ref{fig:quadFig_SEeverLived}): whereas small eccentricities dominate in the population with super-Earths, the values are considerably higher where they are missing.
Planets in $\mathrm{\overline{SE} \cap CJ}$ systems have, in comparison with $\mathrm{SE \cap CJ}$ systems, larger periods.
Many of the super-Earths in $\mathrm{SE \cap CJ}$ systems are on ultra-short orbits of a few days period.
The reason is that such planets are safe from any destructive interaction with outer giants and can thus survive the entire formation and evolution phase.

Similar trends demonstrate instances when the sample is split in metallicity (see upper panels of Fig.~\ref{fig:quadFig_SEeverLived}), which may serve as a proxy for giant planet occurrence~\citep{Johnson2010}. Our multiplicity analysis further shows that we can confirm the observed anti-correlation between cold-Jupiter occurrence and super-Earth multiplicity~\citep[][see Fig.~\ref{fig:multiplicitiesCJnoCJ}]{Zhu2018}.

In summary, these findings suggest that most planetary systems produce super-Earths and where they are missing today, the stability of the system was perturbed by a giant planet. The culprit was typically not a cold Jupiter on a wide orbit, but a dynamically active and massive warm giant.
A note of caution is due here since our model produces, on average, \led{``warmer''}\ giants than those found in the exoplanet demographics.
This might lead to an overestimation of the effect of warm giants.

\subsection{Volatile-poor super-Earths \rev{could be} proxies for giant planets}
\begin{figure*}
        \centering
        \includegraphics[width=\hsize]{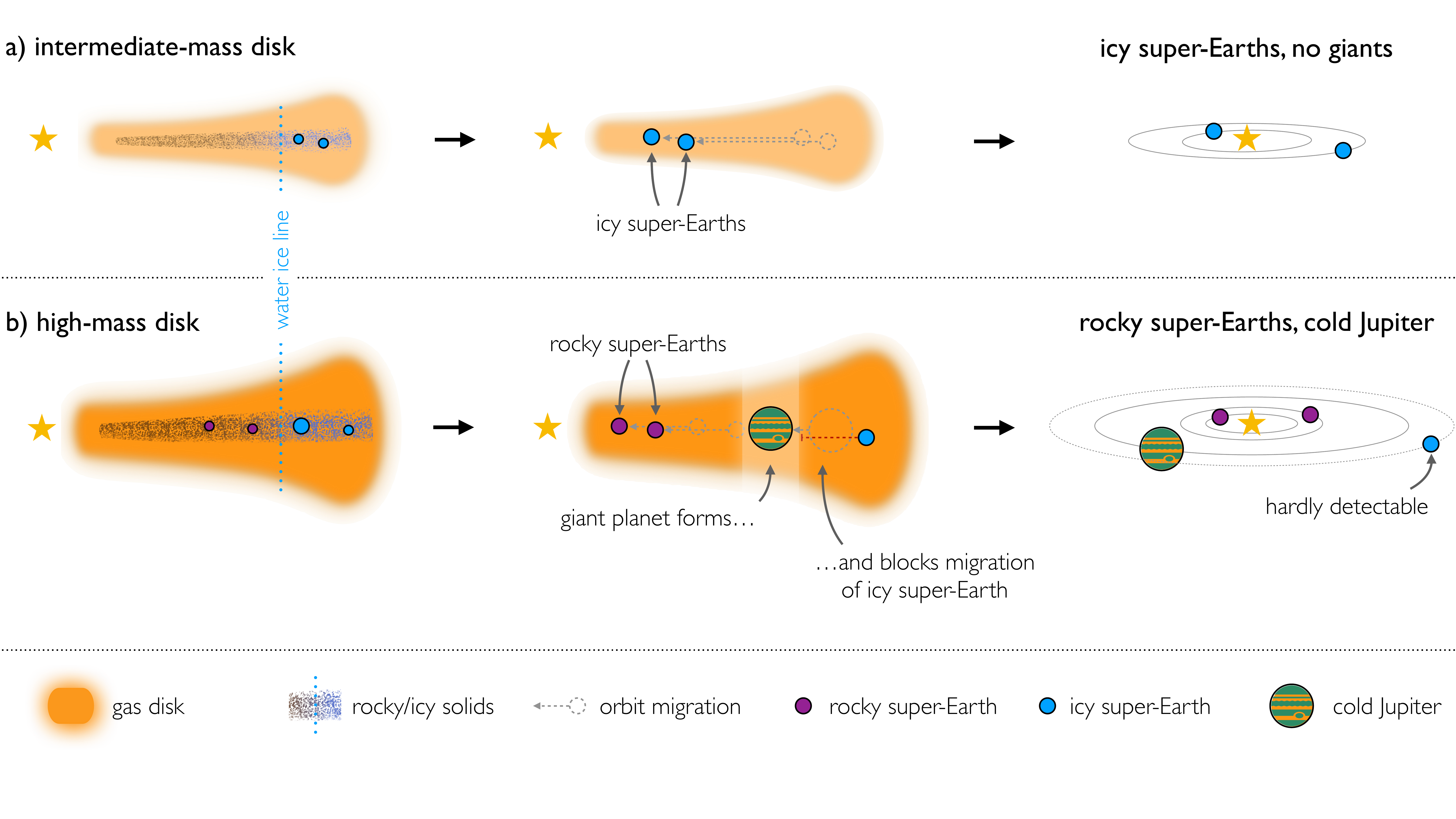}
        \caption{Schematic relations between solid disk mass, core ice, and system architecture:\protect\\
        a) a disk with just enough solid material ($ M_\mathrm{solid}\sim \SI{100}{M_\oplus}$) to grow super-Earth cores will produce them preferably right behind the water ice line. The emerging ice-rich planets remain of too low mass to trigger runaway gas accretion and migrate freely to inner orbits where they can be detected.\protect\\
        b) a more massive disk ($ M_\mathrm{solid}\gtrsim \SI{200}{M_\oplus}$) provides the conditions for giant planet formation, which again happens predominantly just beyond the ice line.
        The gas giant then blocks the migration of ice-rich cores that formed further out. However, the disk allows also for growth of dry super-Earths on closer orbits. In this scenario, the existence of super-Earths with high bulk densities is a proxy for giant planets in the same system.}
        \label{fig:icyRocky_schematics}
\end{figure*}

For planets with both mass and radius measurements, their bulk density can be derived and the compositions of their interiors constrained.
 Most of these planets are expected to be accompanied by additional, often undetected planets~\citep{Zink2019a, Sandford2019}.
Such companions, in particular a hypothetical giant planet, can place strong constraints on the formation history of the system.
Our analysis of planetary compositions in Sect.~\ref{sec:iceMassFractions} imply how under specific conditions the position of a planet in the mass-radius diagram could be used as a proxy for the existence of such an outer gas giant.
A prerequisite for this proposal is a model that is able to produce both ice-rich and dry super-Earths.
This has been proven difficult in the past, as core accretion models typically predict efficient core growth only beyond the water ice line, producing exclusively ice-rich planets~\citep{Izidoro2019}.
Conversely, in the population presented here, super-Earths accompanied by outer gas giants are reduced in volatile species compared to those without a giant companion.
In Fig.~\ref{fig:icyRocky_schematics}, we illustrate schematically the reason for this dichotomy, which is rooted in disparate disk environments:

Disks that produce super-Earths but no cold giants tend to possess intermediate amounts of gas and solids.
Here, most super-Earths start just outside the ice line where the additional reservoir of condensed volatiles provides the most efficient growth of a solid core.
The mass surface density is however too small for the protoplanets to reach the critical masses for giant formation.
They remain in the efficient \rev{Type~I} migration regime and experience strong inward migration, leading to the observed population of icy super-Earths in these systems.

On the other hand, disks that produce both planet types contain large solid and gas reservoirs (compare Fig.~\ref{fig:initialConditionsOfPopulations}), which enables efficient growth of planetary cores to detectable sizes in a large range of orbits.
In such disks, ice-poor \led{super-Earths} can form within the water ice line, while cores that accrete in regions just beyond it typically reach runaway gas accretion and grow to giant planets.
They quickly enter the weaker \rev{Type~II} regime of planet migration and remain cold giants~\citep{Mordasini2018}.
Possible additional planets that formed further out cannot cross the giant's orbit to reach the inner system, which therefore contains only rocky planets.

These findings highlight the strong correlation between the migration history of inner super-Earths and their water content, which is largely determined by the fraction of the accretion phase spent outside of the ice line.
This is also true for pebble accretion models if they assume that inward drifting pebbles lose their water ice once they cross the ice line~\citep{Bitsch2019}.

These differences in the composition of inner super-Earths puts them into different regions in the mass-radius diagram (compare Fig.~\ref{fig:mass-radius_CJ-noCJ}).
While those planets with high ice mass fractions populate regions of larger radii and lower masses, that is, lower density, their rocky counterparts tend to occupy denser regions.
Planets with significant H/He envelopes appear as a distinct group of planets with $\mathrm{R_P} \gtrapprox \SI{3}{R_\oplus}$ in the diagram.
In this regime, the separation is strikingly clear: while super-Earths in systems hosting cold Jupiters have typical masses of several tens of Earth masses, there are practically no gas-rich inner planets with masses below $\sim \SI{10}{M_\oplus}$ in these systems.
The reason is not, as one might suspect, a higher rate of giant impacts that can strip the envelopes of planets with low surface gravity.
The frequency of such events is comparable in both populations.
Instead, the higher core densities of these planets puts them to much higher masses at comparable radii.
While the gaseous envelope contributes the bulk of a planet's radius, the solid core dominates its total mass.

In conclusion, our model makes the testable prediction that volatile-poor super-Earths are more likely to host a long-period giant planet.
Conversely, an inner super-Earth in a system that is also harboring a distant giant planet is likely depleted \led{of} volatiles.

Obviously, this should be interpreted in light of the assumptions we put into our model.
The main simplifications that could influence our results include: the simplified disk chemistry, efficient formation of planetary embryos in the entire disk,
accretion of \SI{300}{\meter}~sized planetesimals, and the generally unsatisfying constraints on planet migration.

Unfortunately, the current sample of exoplanets with known mass and radius is still too small to unequivocally test our hypothesis.
However, we show in Sect.~\ref{sec:iceMassFractions} that the currently available sample of 26 super-Earths with confirmed cold Jupiter-companions matches the predicted bulk densities for such systems much better than the one for the overall synthetic population.
This is especially true for atmosphere-hosting planets, where the observed sample fits only the synthetic super-Earth population with giant companions and not the one without.
A more rigorous benchmark would be a thorough reanalysis of observed planets using raw photometry and RV time series, and including a consistent evaluation of the underlying detection bias.
This is not only beyond the scope of this paper, but will require a larger sample of planets with precise photometric and spectroscopic measurements than is currently available.
With the ongoing \textit{TESS} mission~\citep{Ricker2014} and RV follow-up of its planet candidates, the number of systems for which such data sets exist is constantly increasing.
Thus, statistical tests of the trends we presented here are imminent and will ultimately show if our predictions hold.

\subsection{A negative metallicity correlation for super-Earths in cold-Jupiter hosting systems}\label{sec:negativeMetCorr}
The conditional planet host fractions as a function of host star metallicity (Sect.~\ref{sec:metallicityOccurrences}) reveal an unexpected trend: $\mathrm{P(SE|CJ)}$ correlates negatively with metallicity.
Before we illustrate that this correlation is caused by an increased emergence of \rev{warm} giants in high-metallicity systems, we exclude multiple alternative scenarios:

~\\
\textbf{Multiple giant planets in high-metallicity systems}\\
One factor potentially influencing the relation between metallicity and super-Earth occurrence is the formation of several massive planets per system.
Such efficient formation channels \rev{may be} expected in high-metallicity systems, which emerged from disks containing a large amount of solids.
\rev{However, the multiplicity of giant planets does not vary with metallicity in the synthetic population (compare Fig.~\ref{fig:NgiantsSE_met}).While giant planets are only present from a threshold metallicity upwards, the ratio between systems of different giant planet multiplicity remains the same with increasing [Fe/H] (compare with~\citet{Emsenhuber2020b}).
}

~\\
\textbf{Selection bias induced through our super-Earth definition}\\
Another possibility is that with increasing [Fe/H], we are missing an increasing amount of planets in our statistics because they grow to larger bodies which we do not classify as super-Earths.
Such a deficit would be apparent in the general super-Earth statistics $\mathrm{P(SE)}$, which, \led{however}, shows a positive metallicity correlation, ruling out this assumption.

~\\
\textbf{\rev{Orbital properties of giant planets}}\\
A proposition we want to pursue in more detail is that high-metallicity giants are \rev{on shorter orbits and} dynamically active and \rev{thus} more likely to destroy a population of small planets at short orbital periods.
This would make sense, especially in light of the observed close relation between super-Earth occurrence and disk solid content (compare Sect.~\ref{sec:diskProps}) which, in our model, is tightly correlated with metallicity.
With an increasing amount of solids, the occurrence rate drops just where the first giant planets emerge.
To test this hypothesis, we compared orbital parameters of giants with low- and high-metallicity host stars (Fig.~\ref{fig:CJdistributions}) and found that the high-metallicity sample extends to lower periods.
Their mass distribution is inconspicuous, but their eccentricities are slightly enhanced.

\begin{figure}
        \centering
        \includegraphics[width=\hsize]{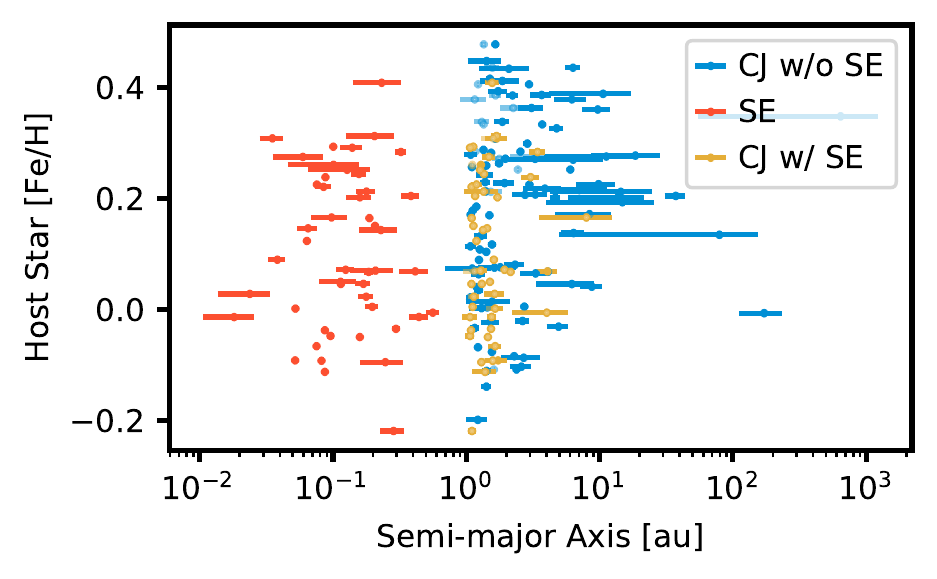}
        \caption{Orbital range of outermost super-Earth and innermost cold Jupiter \rev{plotted by} host star metallicity. For each system, we plot the orbital range (periapsis to apoapsis) of the outermost super-Earth \rev{(red)} and innermost cold Jupiter \rev{(yellow).
        Cold Jupiters in systems without super-Earths are shown in blue, and }
        light markers correspond to giant planets that did not survive. Cold Jupiter systems with host star metallicities greater than ${\sim 0.3}$ typically harbor no super-Earths, and high-metallicity giants without super-Earths are often on eccentric orbits.}
        \label{fig:outerSEinnerCJmet}
\end{figure}

We are especially interested in giants with small and intermediate orbital distances since these planets are most prone to disturbing inner super-Earths.
Even though, in our model, the planets that endanger these systems are mainly warm Jupiters, in the following, we briefly discuss the planets fulfilling our criteria for a  coldJupiter (see Table~\ref{tab:planetTypes}).
Figure~\ref{fig:outerSEinnerCJmet} relates, for each \rev{pair of inner super-Earth and cold Jupiter, the} semi-major axes of the innermost giant \rev{(yellow)} and of the outermost super-Earth \rev{(red)} with the metallicity of their host star.
\rev{For comparison, we plot cold Jupiters that do not have inner super-Earth companions in blue.}
Horizontal lines denote the full orbital ranges from periapsis to apoapsis and light markers correspond to giant planets that formed but have not survived.
Between the outermost super-Earths and their innermost cold Jupiter companion, a ``safety gap'' of $\sim \SI{1/2}{au}$ emerges.
Almost no giants with host star metallicities greater than $[\mathrm{Fe/H}] \approx 0.3$ have super-Earth companions and the aforementioned positive correlation between host star metallicity and cold Jupiter eccentricity is apparent. The highest eccentricities belong to giant planets in systems that lost their super-Earths.

Taken together, these results indicate that inner super-Earth systems are more likely to be destroyed in a high-metallicity environment, where \rev{there are warmer and dynamically more active } giant planets  that can disrupt them.

\section{Conclusions}\label{sec:conclusions}

In the NGPPS series of articles, we present a population synthesis of planetary systems from the Bern \textit{Generation 3} global planet formation and evolution model.
In this, the third paper, we compared this population to observed exoplanets around Solar-type stars, focusing on the relation of close-in super-Earths and far-out giant planets (``cold Jupiters'').
Our results can be summarized as follows:

\begin{enumerate}
	\item Our synthetic planet population shows a positive intra-system correlation between the occurrences of inner super-Earths and cold Jupiters, albeit weaker than previously \led{proposed}.
        The reduction is attributed to warm giant planets that frequently disrupt inner systems of super-Earths. This discrepancy might hint to an overestimation of the migration efficiency of giant planets.
        \rev{We showed that the correlation is sensitive on the choice of mass and period limits that defines the sample of inner and outer planets.}
        \item We find a difference in the bulk composition of inner super-Earths with and without cold Jupiters.
        High-density super-Earths point to the existence of outer giant planets in the same system.
    Conversely, a present cold Jupiter gives rise to rocky, volatile-depleted inner super-Earths.
        Birth environments that produce such dry planet cores in the inner system are also favorable for the formation of outer giants, which obstruct inward migration of icy planets that form on distant orbits. This predicted correlation can be tested observationally.
        \rev{\item It is the result of a general link between the initial reservoir of solids and final system architecture: low-mass solid disks tend to produce only super-Earths but no giant planets.
        Intermediate-mass disks may produce both super-Earths and cold Jupiters.        High-mass disks lead to the destruction of super-Earths and only giants remain.
        }
        \item Inner super-Earths initially form in nearly all systems that host an outer giant. Where they are missing today, the inner system was dynamically excited by giant planets on intermediate orbits, leading to the destruction of super-Earths.
        \item The key parameter for the formation of both cold Jupiters and super-Earths is the solid content of the protoplanetary disk. With increasing initial solid mass, super-Earth occurrence rises steeply but drops for disks that are massive enough to form giant planets.%
        \item Outer giants reduce the multiplicity of small inner planets. In line with the tentative observational evidence \citep{Zhu2018}, the number of super-Earths that survive the entire formation and evolution phase \rev{is reduced} where cold Jupiters occur.
        \item High-metallicity giant planet hosts are less likely to harbor inner super-Earths. Planetary systems around stars with high metallicity frequently contain \rev{warm and dynamically active} giant planets that can disrupt inner planetary systems.
\end{enumerate}

\begin{acknowledgements}
We thank Bertram Bitsch, Miriam Keppler, and Christian Lenz for fruitful discussions and stimulating comments.
The authors are grateful for very constructive feedback by an anonymous referee.
C.M. acknowledges the support from the Swiss National Science Foundation under grant BSSGI0$\_$155816 ``PlanetsInTime''. Parts of this work have been carried out within the framework of the NCCR PlanetS supported by the Swiss National Science Foundation.
This work was supported by the DFG Research Unit FOR2544 “Blue Planets around Red Stars”, project no. RE 2694/4-1.
This research has made use of the NASA Exoplanet Archive, which is operated by the California Institute of Technology, under contract with the National Aeronautics and Space Administration under the Exoplanet Exploration Program.

\end{acknowledgements}

\bibliographystyle{aa} %
\bibliography{PhD,addLit} %

\begin{appendix}

\section{\rev{Monte Carlo} parameters}\label{sec:initialConditions}
\rev{In this appendix, we present the motivations behind the chosen distributions for our Monte Carlo parameters.}

~\\
\textbf{Initial gas disk mass $M_\mathrm{gas}$}\\
The gas content largely governs the mass of a protoplanetary disk~\citep[][however, see \citet{Miotello2017}]{Ansdell2016b}.
While the intrinsic distribution of gas disk masses is poorly constrained, observations point to masses ranging between \SI{0.1}{\percent} and \SI{10}{\percent} of the stellar host mass~\citep{Andrews2010, Manara2016}.
To compute the gas mass of our computational disk $M_\mathrm{gas}$,  we drew $\log \frac{M_\mathrm{gas}}{M_\odot}$ from a normal distribution $\mathcal{N}(\mu=-1.49,\,\sigma^2=0.123)$~\citep{Tychoniec2018}.

In combination with the disk radius $R_\mathrm{cut,g}$ (see below), the resulting initial gas surface density at a reference radius of \SI{5.2}{au},  $\Sigma_0$, varies log-normally with a median of \SI{132}{\gram\per\centi\meter\squared}.%
Note that the range of $M_\mathrm{gas}$ was cut to avoid both extremely massive and very low-mass disks.

~\\
\textbf{Dust-to-gas ratio $\zeta_\mathrm{d,g}$}\\
\label{sec:hostMet}
We assumed that the bulk metallicity of the disk $[\mathrm{M}/\mathrm{H}]_\mathrm{disk}$ equals the heavy-element abundances of the protostar and modeled it as a normal distribution with a mean $\mu = -0.03$  and a standard deviation $\sigma = 0.20$ that follows the dispersion of stellar metallicities in the Solar neighborhood~\citep[][compare Fig.~\ref{fig:hostStarMet}]{Santos2005}.

\begin{figure*}
        \centering
        \includegraphics[width=\hsize]{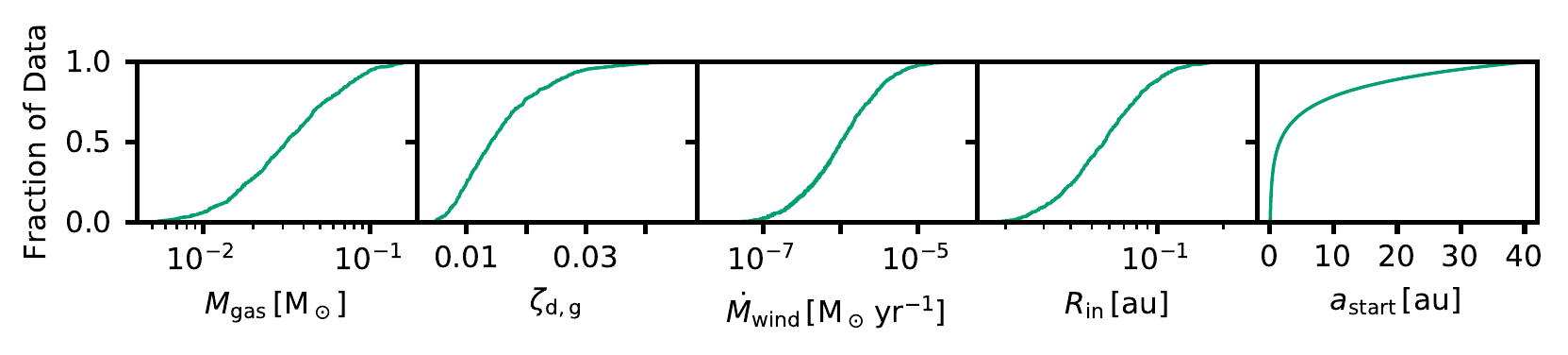}
        \caption{\rev{Cumulative distribution} of Monte Carlo parameters. The disk initial conditions, which were drawn randomly from these distributions for each simulation, comprise the initial gas disk mass $M_\mathrm{gas}$, the dust-to-gas ratio $\zeta_\mathrm{d,g}$, the photoevaporative mass loss rate $\dot{M}_\mathrm{wind}$, the inner disk edge $R_\mathrm{in}$, and the starting locations of the planetary embryos $a_\mathrm{start}$.}
        \label{fig:initialConditions}
\end{figure*}

\begin{figure}
        \centering
        \includegraphics[width=\hsize]{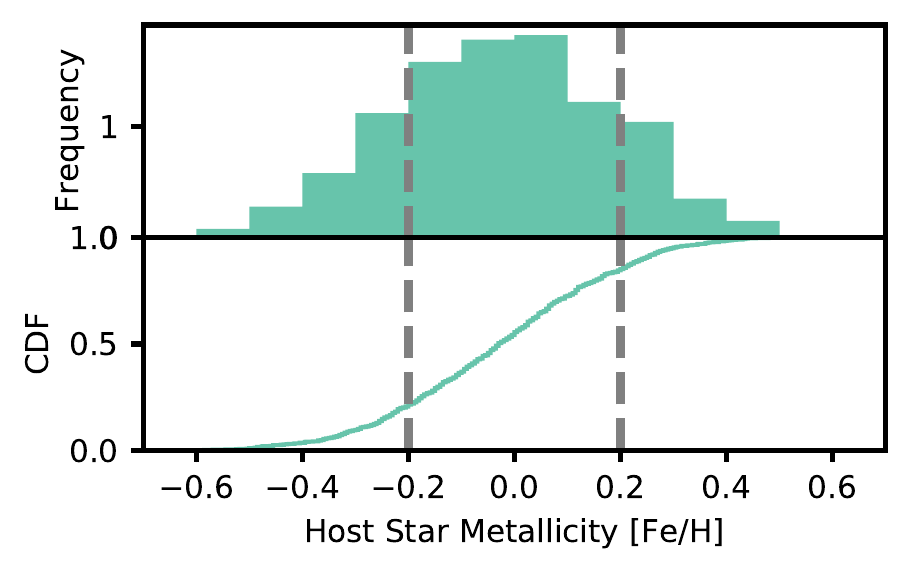}
        \caption{Distribution of host star metallicities. \textit{Upper panel:} Frequency normalized to one. \textit{Lower panel:} Empirical distribution function showing the fraction of host stars that are less than or equal to the specified metallicity. Dashed lines denote the borders of the metallicity bins used in this study.}
        \label{fig:hostStarMet}
\end{figure}

$[\mathrm{M}/\mathrm{H}]_\mathrm{disk}$ is then readily converted to the dust-to-gas ratio of the disk via
\begin{equation}
\zeta_\mathrm{d,g} = Z_0 \cdot 10^{[\mathrm{M}/\mathrm{H}]_\mathrm{disk}},
\end{equation}
where $Z_0 = 0.0149$~\citep{Lodders2003}.
Its distribution is shown in Fig.~\ref{fig:initialConditions}.
In combination with the disk's gas content, the dust-to-gas ratio determines the amounts of solids available for forming planets.

~\\
\textbf{Photoevaporative mass loss rate $\dot{M}_\mathrm{wind}$}\\
A constraining factor for a system's ability to build planets is the lifetime of its protoplanetary disk.
Large IR surveys of young stellar clusters enabled to identify the fraction of stars with disks as a function of their age, indicating a disk lifetime of a few \SI{}{\mega yr}~\citep{Haisch2001, Mamajek2009, Fedele2010}.
It is believed that photoevaporation from high-energy photons, in combination with viscous accretion, is the main mechanism for the rapid dispersal of disks at the end of their lifetimes~\citep{Clarke2001, Owen2012}.
We parameterized this effect with a mass loss rate $\dot{M}_\mathrm{wind}$, which we varied such that the distribution of synthetic disk lifetimes is similar to the observed distribution in \citet{Fedele2010}.
The parameter was normalized to a hypothetical disk extending to \SI{1000}{au} and thus does not equal the absolute mass loss rates of our disks.
~\\
\textbf{Inner disk edge $R_\mathrm{in}$}\\
The physical motivation for an inner edge of the gas disk is the development of a magnetospheric cavity~\citep{Bouvier2007}, which is thought to extend to the corotation radius, i.e. the location where the angular velocity of the stellar magnetic field and of the orbiting gas are equal~\citep[e.g.,][]{Guenther2013}.
For the numerical disk, we adopted the orbit radius corresponding to the rotation period of its host star for $R_\mathrm{in}$.
We drew these periods from a distribution based on recent measurements in the young stellar cluster NGC 2264, which has an estimated age of \SI{3}{\mega yr}~\citep{Venuti2017}.
The resulting log-normal distribution has a mean period of \SI{4.74}{\day} \rev{and a standard deviation of \SI{0.31}{dex}}.

~\\
\textbf{Disk radius $R_\mathrm{cut,g}$}\\
The sizes of protoplanetary disks are both constrained by observations~\citep[e.g.,][]{Andrews2010, Andrews2018, Ansdell2018} and by analytically solving the viscous accretion disk problem~\citep{Lynden-Bell1974}.
In the model, the radial extent of the gas disk $R_\mathrm{cut,g}$ is not an independent Monte Carlo variable but we computed it from $M_\mathrm{gas}$ using a scaling relation derived from millimeter continuum emission sizes~\citep{Andrews2018}.
Our disk radii range from roughly \SI{20}{au} to \SI{150}{au}.

The solid material is represented by a continuous disk of solids.
To take into account spatial concentration due to inward drift of solid material~\citep{Weidenschilling1977}, its size is initially half of that of the gas disk.
This choice of scaling is motivated by millimeter observations of the Lupus star-forming region, which suggest a factor of two difference between gas and dust disks~\citep{Ansdell2018}.

~\\
\textbf{Starting location of planet embryos $a_\mathrm{start}$}\\
For each system, we inserted 100 planetary seeds with a starting mass of $\SI{0.01}{M_\oplus}$ into the disk.
The initial location of these embryos were randomly drawn from a log-uniform distribution in semi-major axis.
This follows N-body simulations of planetary embryos that found oligarchs spaced by a few Hill spheres, that is, their separations are proportional to their orbital distance \citep{Kokubo2000}.
The roughly Moon-mass seeds were distributed from the inner edge of the disc $R_\mathrm{in}$ up to \SI{40}{au} with the additional constraint that no embryo was placed closer than 10 Hill radii to another.

~\\
For a more detailed description of the observational and theoretical grounds of these Monte Carlo parameters, we refer to \citet{Mordasini2009a, Mordasini2009}, \citet{Mordasini2012a} and \papertwo.

Table~\ref{tab:modelParams} shows the distributions of Monte Carlo parameters chosen for our simulations, as well as additional parameters we kept fixed.
The distributions of initial conditions for the different populations discussed in the main text are listed in Table~\ref{tab:initialConditionsOfPopulations}.

\begin{table*}
        \caption{Model Parameters}             %
        \label{tab:modelParams}
        \centering                          %
        \begin{tabular}{r c c c}        %
                \hline\hline                 %
                Parameter & Symbol & Distribution & Range or Median$\substack{+84\% \\ -16\%}$\\
                \hline                        %
                \textbf{Fixed Parameters}& & & \\
                Stellar Mass & & -- & \SI{1}{M_\odot}\\
                Disk Viscosity & $\alpha$ & -- & $\SI{2e-3}{}$ \\
                Power Law Index (Gas) & $\beta_\mathrm{g}$ & -- & $0.9$\\
                Power Law Index (Solids) & $\beta_\mathrm{s}$ & -- & $1.5$\\

                Radius of Planetesimals & & -- & \SI{300}{\meter} \\
                Number of Planet Seeds & & -- & $100$ \\
                Mass of Planet Seeds & & -- & \SI{0.01}{M_\oplus} \\
                \hline
                \textbf{Monte Carlo Parameters}& & & \\
                Host Star Metallicity& [Fe/H]& normal & $-0.03 \pm 0.20$\\
                Initial Gas Surface Density at \SI{5.2}{au} & $\Sigma_0$& log-normal & $132 \substack{+37 \\ -27}\,\SI{}{\gram\per\centi\meter\squared}$\\
                Inner Disk Radius & $R_\mathrm{in}$ & log-normal & $ 4.74 \substack{+4.94\\ -2.42} \,\SI{}{\day}$ \\
                Gas Disk Cutoff Radius & $R_\mathrm{cut,g}$& log-normal & $ 56 \substack{+36 \\ -21}\,\SI{}{au}$\\
                Solid Disk Cutoff Radius & $R_\mathrm{cut,s}$& log-normal & $R_\mathrm{cut,g}/2$ \\
                Photoevaporation Efficiency & $\dot{M}_\mathrm{wind}$ & log-normal & $ (1.0 \substack{+2.2 \\ -0.7})\times 10^{-6}\,\SI{}{M_\odot/yr}$ \\
                Starting Position of Planet Seeds& $a_\mathrm{start}$b & uniform in $\log a$ & $R_\mathrm{in} \,\mathrm{to} \,40\,\mathrm{au}$\\
                \hline
                \textbf{Derived Parameters}& & & \\
                Initial Gas Disk Mass & $M_\mathrm{gas}$ &  log-normal & $0.03 \substack{+0.04 \\ -0.02}$ \SI{}{M_\odot}\\
                Initial Solid Disk Mass& $M_\mathrm{solid}$ & $\sim$ log-normal & $95 \substack{+147 \\ -55}$\SI{}{M_\oplus}\\
                Dust-to-gas Ratio  & $\zeta_\mathrm{d,g}$ & log-normal & $0.02 \substack{+0.01 \\ -0.01}$\\
                Disk Dispersal Time& $t_\mathrm{disk}$& -- & $(3.2 \substack{+1.9 \\ -1.0})\times 10^{6}\,\SI{}{yr}$ \\
                \hline
        \end{tabular}
        \tablefoot{Upper panel: Initial conditions that are fixed for each simulation. Middle panel: Monte Carlo parameters that are drawn randomly. Lower panel: Quantities that are derived from or controlled by other parameters. Upper and lower limits denote 84th and 16th percentiles, respectively.}
\end{table*}

\begin{table*}
        \caption{Initial Conditions of different populations}             %
        \label{tab:initialConditionsOfPopulations}
        \centering                          %
        \begin{tabular}{r c c c c c c}        %
                \hline\hline                 %

population & [Fe/H] & $M_\mathrm{solid}\, [\mathrm{M_\oplus}]$ & $M_\mathrm{gas}\, [\mathrm{M_\odot}]$ & $R_\mathrm{cut,g}\, [\mathrm{au}]$ & $t_\mathrm{disk}\, [\mathrm{Myr}]$ & $a_\mathrm{start}\, [\mathrm{au}]$\\
\hline
CJ  & 0.13 $\substack{+0.15\\ -0.17}$ & 290 $\substack{+127\\ -56}$ & 0.07 $\substack{+0.03\\ -0.03}$ & 95 $\substack{+21\\ -24}$ & 3.35 $\substack{+2.26\\ -1.24}$ & 22 $\substack{+12\\ -17}$\\
SE  & 0.05 $\substack{+0.19\\ -0.19}$ & 156 $\substack{+70\\ -47}$ & 0.04 $\substack{+0.03\\ -0.02}$ & 69 $\substack{+25\\ -18}$ & 3.90 $\substack{+1.90\\ -1.28}$ & 16 $\substack{+14\\ -11}$\\
$\mathrm{SE \cap CJ}$  & 0.07 $\substack{+0.19\\ -0.12}$ & 283 $\substack{+84\\ -50}$ & 0.07 $\substack{+0.03\\ -0.03}$ & 95 $\substack{+21\\ -24}$ & 3.35 $\substack{+1.96\\ -1.38}$ & 20 $\substack{+13\\ -15}$\\
$\mathrm{\overline{SE} \cap \overline{CJ}}$  & -0.11 $\substack{+0.20\\ -0.19}$ & 54 $\substack{+41\\ -29}$ & 0.02 $\substack{+0.02\\ -0.01}$ & 42 $\substack{+22\\ -14}$ & 2.84 $\substack{+1.14\\ -0.81}$ & 11 $\substack{+16\\ -9}$\\
$\mathrm{SE \cap \overline{CJ}}$  & 0.05 $\substack{+0.19\\ -0.20}$ & 145 $\substack{+51\\ -39}$ & 0.04 $\substack{+0.03\\ -0.01}$ & 66 $\substack{+23\\ -16}$ & 4.07 $\substack{+1.74\\ -1.43}$ & 15 $\substack{+14\\ -10}$\\
$\mathrm{\overline{SE} \cap CJ}$  & 0.17 $\substack{+0.12\\ -0.18}$ & 302 $\substack{+129\\ -66}$ & 0.07 $\substack{+0.04\\ -0.03}$ & 95 $\substack{+26\\ -25}$ & 3.41 $\substack{+2.29\\ -1.29}$ & 24 $\substack{+11\\ -18}$\\
all  & -0.03 $\substack{+0.20\\ -0.20}$ & 99 $\substack{+90\\ -50}$ & 0.03 $\substack{+0.03\\ -0.02}$ & 56 $\substack{+30\\ -19}$ & 3.40 $\substack{+1.85\\ -0.98}$ & 5 $\substack{+5\\ -4}$\\
initial  & -0.03 $\substack{+0.22\\ -0.21}$ & 95 $\substack{+147\\ -55}$ & 0.03 $\substack{+0.04\\ -0.02}$ & 56 $\substack{+36\\ -21}$ & 3.23 $\substack{+1.90\\ -0.98}$ & 2 $\substack{+13\\ -1}$\\

                \hline
        \end{tabular}
        \tablefoot{Initial Conditions for different populations. For each parameter, we quote its median for all combinations of SE and CJ, plus for the entire population of survived planets. Upper and lower limits denote 84th and 16th percentiles, respectively. Compare Fig.~\ref{fig:initialConditionsOfPopulations} for a visual representation of the data.}
\end{table*}

\newpage
\section{\rev{Planet radii in the synthetic population}}
\begin{figure}
        \centering
        \includegraphics[width=\hsize]{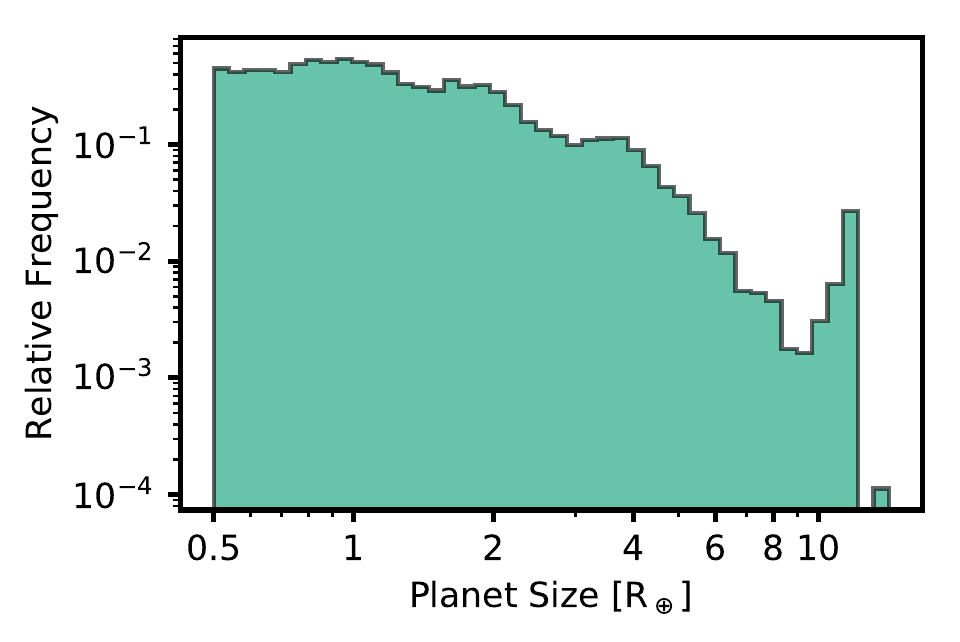}
        \caption{Distribution of planet radii in the synthetic population. We exclude planets smaller than \SI{0.5}{R_\oplus} and with periods beyond \SI{3000}{\day}.
        The radius frequencies follow a distinct bimodal distribution with the bulk at its low-size end.
        }
        \label{fig:radiusHist}
\end{figure}

Figure~\ref{fig:radiusHist} reveals a bimodal structure in the radius distribution of our synthetic population: most planets are terrestrial or super-Earth-sized, but an additional, shallower local maximum close to $1\, \mathrm{R_{Jup}}$ exists in the radius distribution.
This bimodality separates giant planets that experienced runaway gas accretion from planets that did not and was seen already in earlier generations of population synthesis models~\citep{Mordasini2012a}.

It is noticeable that even though planets with masses far beyond $1\, \mathrm{M_{Jup}}$ (see Fig.~\ref{fig:occMapMasses}) occur in our population, the radius distribution shows a sharp cutoff at $\sim 12\, \mathrm{R_{Earth}}$.
This feature also appears in the observed exoplanet population~\citep[e.g.,][]{Chen2016} %
and is explained by electron degeneracy in the interior of giant planets~\citep[e.g.,][]{Chabrier2009}.
Close to Jupiter mass, the polytropic index  $n \sim 1$ in the equation of state and the radius is independent of the mass.
This leads to a wide range of planet masses populating a narrow region in planet radii.
In the synthetic population shown here, this effect is enhanced since we assume the same atmospheric opacity for all planets  and show all planets at the same age of \SI{5}{\giga yr}~\citep{Mordasini2012a}.

A prominent feature in planetary radius-period diagrams is a depleted region separating small super-Earths from larger sub-Neptunes.
This ``photoevaporation valley'' was predicted by formation and evolution models~\citep{Jin2014} and later confirmed observationally~\citep{Fulton2017, Hsu2018}.
While originally explained by photoevaporation, alternative processes have also been hypothesized to produce the pattern.
Debated mechanisms include core-powered mass loss, where the core's internal luminosity removes the planetary atmosphere~\citep{Ginzburg2016a, Ginzburg2018, Gupta2019}; different formation pathways of planets above and below the gap~\citep{Zeng2019}; and planetesimal impacts~\citep[e.g.,][]{Liu2015, Wyatt2019a}.

Our synthetic population reproduces the radius valley at most in an attenuated form.
The reason for this lies presumably in our simplified treatment of collisional envelope stripping, where we add the impact luminosity of a collision event to the intrinsic planetary luminosity~\rev{\citep{Emsenhuber2020}.}
In contrast to photoevaporation from high-energy photons from the star, this mechanism not only affects the innermost region of the system but also the envelopes of planets further out.
Also, more massive planets suffer from atmospheric loss than it is the case with photoevaporation alone.
Both effects fill up the radius valley.
When the luminosity from impacts is neglected in our model and atmospheres are stripped only by photoevaporation, a significant radius valley emerges (compare \citet{Jin2018}).
At least two possible shortcomings of our current prescription would be plausible to explain the observed mismatch with the empirical radius distribution: There might be less collision events than assumed, or they do not remove atmospheres as efficiently as modeled.
Further studies will aim at distinguishing these possibilities as well as the contributions of different atmosphere-depletion mechanisms.

\subsection{Relation between metallicity and planet radius}
\begin{figure}
        \centering
        \includegraphics[width=\hsize]{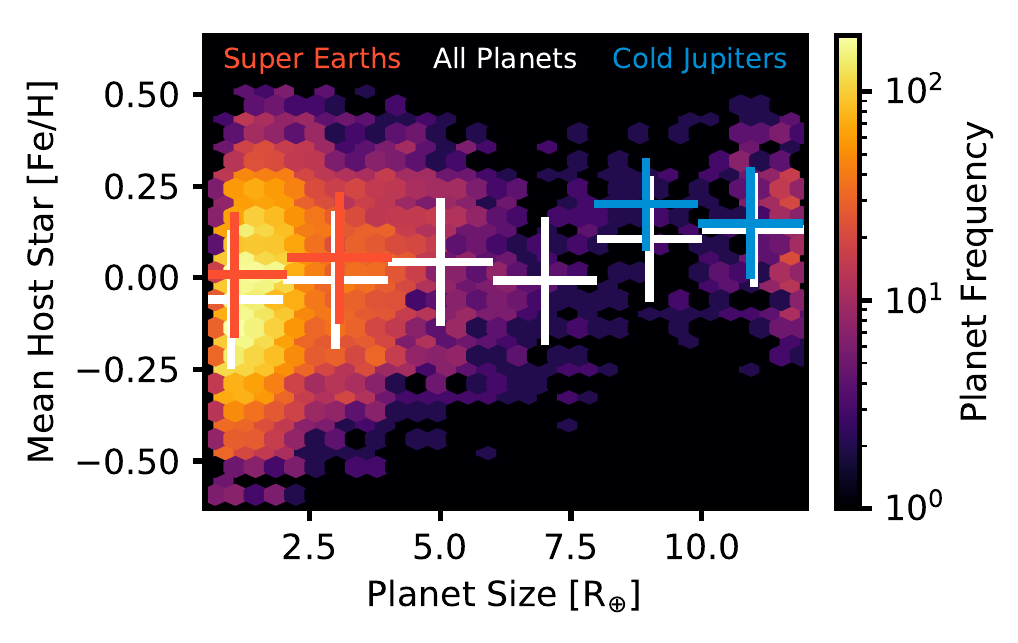}
        \caption{Dependence of host star metallicity on planet radius. For each radius bin, the cross denotes the mean [Fe/H] and the vertical bar is its standard deviation. The markers are slightly offset horizontally for clarity. Overall, there is a positive metallicity trend with planet size. Hosts of super-Earths show marginally higher metallicity with respect to the overall sample. Cold Jupiters are enhanced in metallicity.
}
        \label{fig:radius-meanMet}
\end{figure}
The frequency histogram in Fig.~\ref{fig:radius-meanMet} illustrates the dependencies between host star metallicity and planet radius for different planet types.
It includes all planets with radii between \SI{0.5}{R_\oplus} and \SI{12}{R_\oplus}.
The planet frequency in the metallicity-radius plane reveals a clear positive correlation of gas giant occurrence and stellar metallicity, in agreement with the well-established correlation in the observed exoplanet population~\rev{\citep{Santos2004, Fischer2005, Johnson2010, Buchhave2018}.} Our occurrence density confirms the observed paucity of large planets with sub-Solar metallicity~\citep{Petigura2018}. Small terrestrial planets populate a wide range of metallicities and their host stars are not enhanced in [Fe/H].

Overplotted is a statistics binned in radius for different planet types; vertical bars denote the standard deviations in each bin.
When all planets are considered, their average host star metallicities are consistent with the observed trends in \citet{Buchhave2014} and \citet{Narang2018}.

While all types of giant planets are enhanced in metallicity, cold Jupiters between \SIrange{8}{10}{R_\oplus} are more metal-rich than their siblings in the same size range.
This difference is not related to their orbital distance but due to our classification based on their mass (compare Table~\ref{tab:planetTypes}), which excludes large planets with $M_\mathrm{P} < \SI{95}{M_\oplus}$ from being classified as cold Jupiters.
On average, our cold Jupiters have a higher bulk density and thus contain more solids compared to the entirety of planets in this radius range.

 For super-Earths, we find only a weak positive trend with planet radius. Also, the metallicities of stars harboring these planets are not significantly enhanced compared to the full population. This is consistent with findings of \citet{Sousa2018} who report indistinguishable metallicity distributions of Solar neighborhood stars and stars hosting low-mass planets, respectively.

\newpage
\section{Expected detections of super-Earths}\label{sec:Barbato}
The high conditional probabilities P(SE|CJ) found in \citet{Zhu2018} and \citet{Bryan2019} seem to be in disagreement with \citet{Barbato2018}, who found no inner low-mass planets in a sample of 20 Solar-type stars hosting long-period giants. We note that their definition of super-Earths differs from the one in \citet{Zhu2018} and the survey is not complete in the respective mass-period range. Their super-Earths have $M \sin i$ between $10$ and \SI{30}{M_\oplus} and reside on orbits with periods less than \SI{150}{\day}. For such planets, they ``conservatively'' assume a detection sensitivity $\mathrm{P_{detect}} = 0.5$.
To mirror the survey in a numerical experiment, we repeatedly drew a pseudo-random number $x \in [0.0, 1.0)$ and counted a ``detection'' if:
\begin{equation}
        x < \mathrm{P_{detect}}\cdot \mathrm{P(SE|CJ)},
\end{equation}
where $\mathrm{P_{detect}}$ is the probability to detect an existing super-Earth system and $\mathrm{P(SE|CJ)}$ is the fraction of systems hosting super-Earths in cold Jupiter-hosting systems. Each round of 20 iterations represents a realization of the survey with corresponding $N$ detections. We repeated this experiment  \SI{10000}{} times to obtain a probability for each $N$.

\begin{figure}
        \centering
        \includegraphics[width=\hsize]{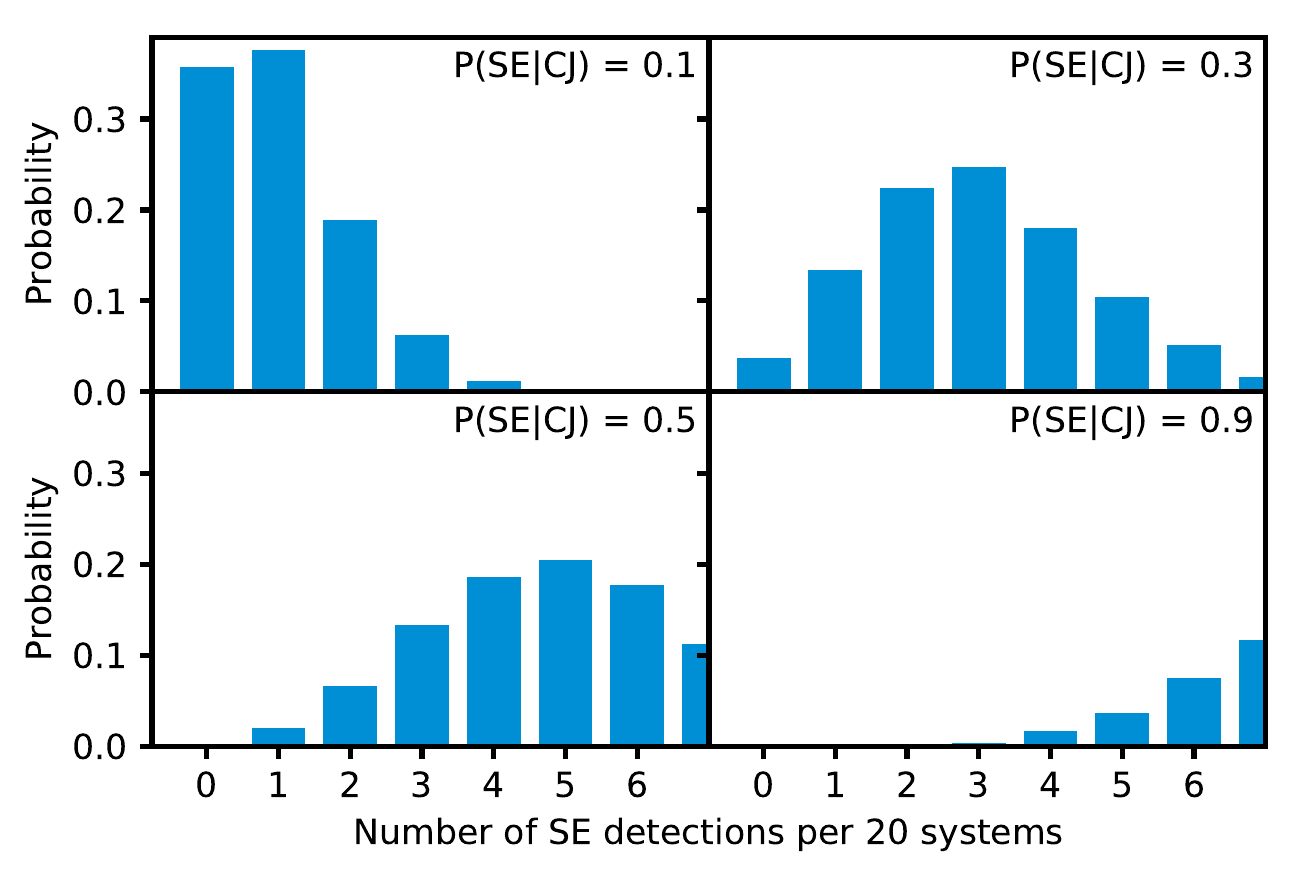}
        \caption{Probability of finding different numbers of super-Earths in a survey of 20 target stars. Each panel assumes a different conditional super-Earth probability ranging from $\mathrm{P(SE|CJ)} = 0.1$ to $0.9$ and we adopt a survey sensitivity for this planet type of $\mathrm{P_{detect}} = 0.5$. The probability to find zero planets approaches zero for P(SE|CJ) greater than $\sim 0.5$.}
        \label{fig:Barbato}
\end{figure}

Figure~\ref{fig:Barbato} shows the detection probabilities of such a survey for $\mathrm{P_{detect}} = 0.5$ and four different occurrence probabilities. For a very low value $\mathrm{P(SE|CJ)} = 0.1$, the probability to find zero super-Earths is as high as $\mathrm{P} (N=0) = 0.35$. If, on the other hand, the probability is $\mathrm{P(SE|CJ)} = 0.3$, this value drops to $\mathrm{P} (N=0) = 0.04$. For probabilities of $\mathrm{P(SE|CJ)} = 0.5$ or higher, $\mathrm{P} (N=0)$ approaches zero. It is extremely unlikely to find zero planets in 20 systems if $\mathrm{P(SE|CJ)} \gtrsim 0.5$ and $\mathrm{P_{detect}} = 0.5$.

Regardless of the different super-Earth definitions in \citet{Zhu2018} and \citet{Barbato2018}, the latter do not detect any sub-giant planets in their sample systems. If we adopt the numbers reported by \citet{Zhu2018} for such planets, their conditional super-Earth probability is $\mathrm{P(SE|CJ)} = 0.9$, and the average sensitivity of the survey must be as low as $0.15$ to obtain a probability of \SI{5}{\percent} for their null result. It is thus difficult to reconcile the results of \citet{Zhu2018}, \citet{Bryan2016}, and \citet{Herman2019} with the one presented in \citet{Barbato2018}.
\rev{Recently, \citet{Barbato2020} stated an impaired sensitivity for their survey, which could explain the non-detection.}

\newpage
\section{Time evolution of individual systems}
\begin{figure*}
        \centering
        \includegraphics[width=\hsize]{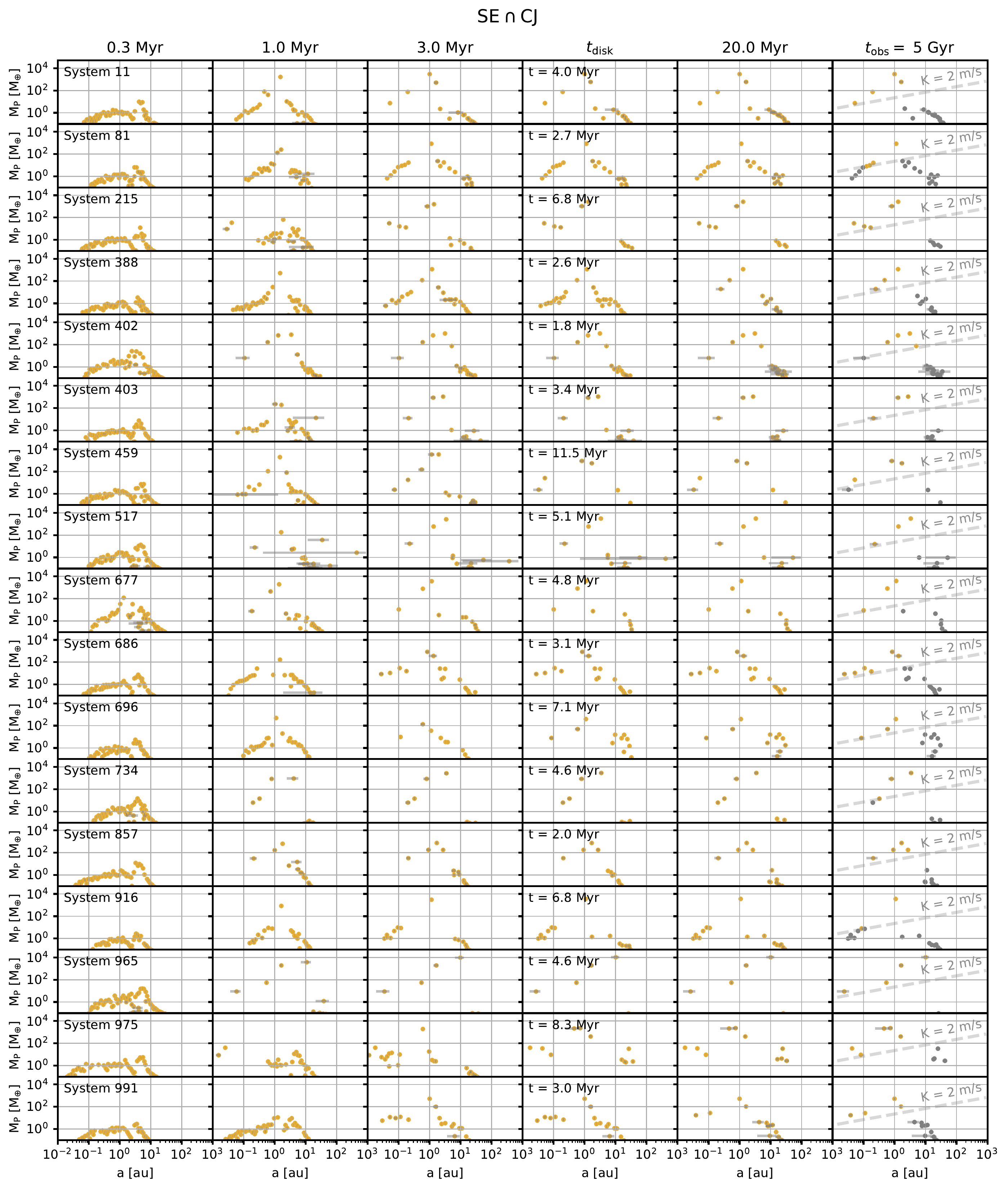}
        \caption{Time evolution of systems with cold Jupiters and super-Earths. For a number of randomly sampled systems, we show the mass-semi-major axis relation of the planets at six different times, where $\mathrm{t_{disk}}$ is the disk dispersal time and $\SI{20}{\mega yr}$ is the integration time of the N-body module. In the last column, unobservable planets are grayed out and the dashed line indicates the detection limit of $\SI{2}{\meter\per\second}$. Horizontal gray lines visualize the orbital range of eccentric planets.
        }
        \label{fig:sysEvo_SEandCJ}
\end{figure*}

\begin{figure*}
        \centering
        \includegraphics[width=\hsize]{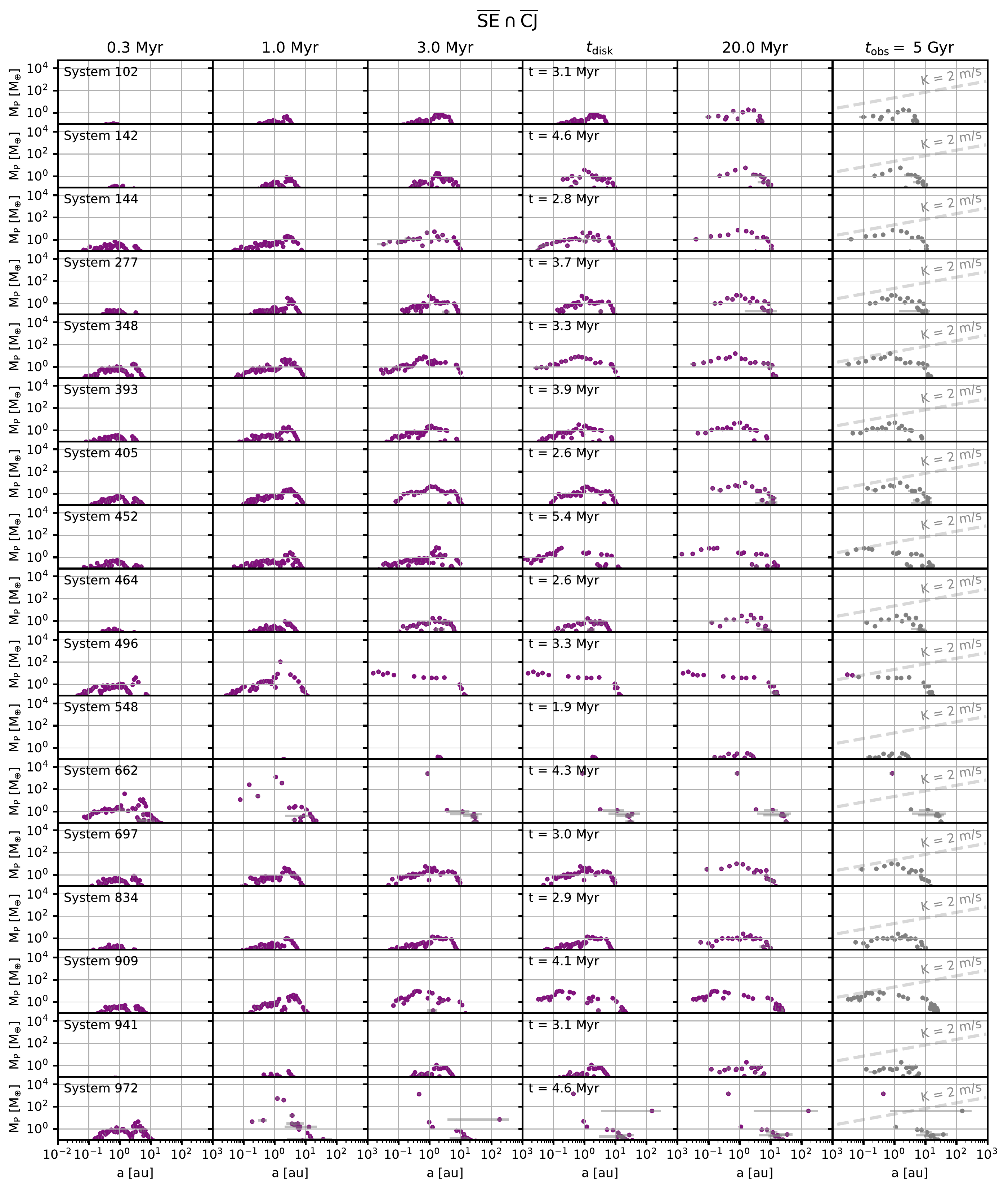}
        \caption{Same as Fig.~\ref{fig:sysEvo_SEandCJ}, but for systems containing neither super-Earths nor cold Jupiters.
        }
        \label{fig:sysEvo_noSEandnoCJ}
\end{figure*}

\begin{figure*}
        \centering
        \includegraphics[width=\hsize]{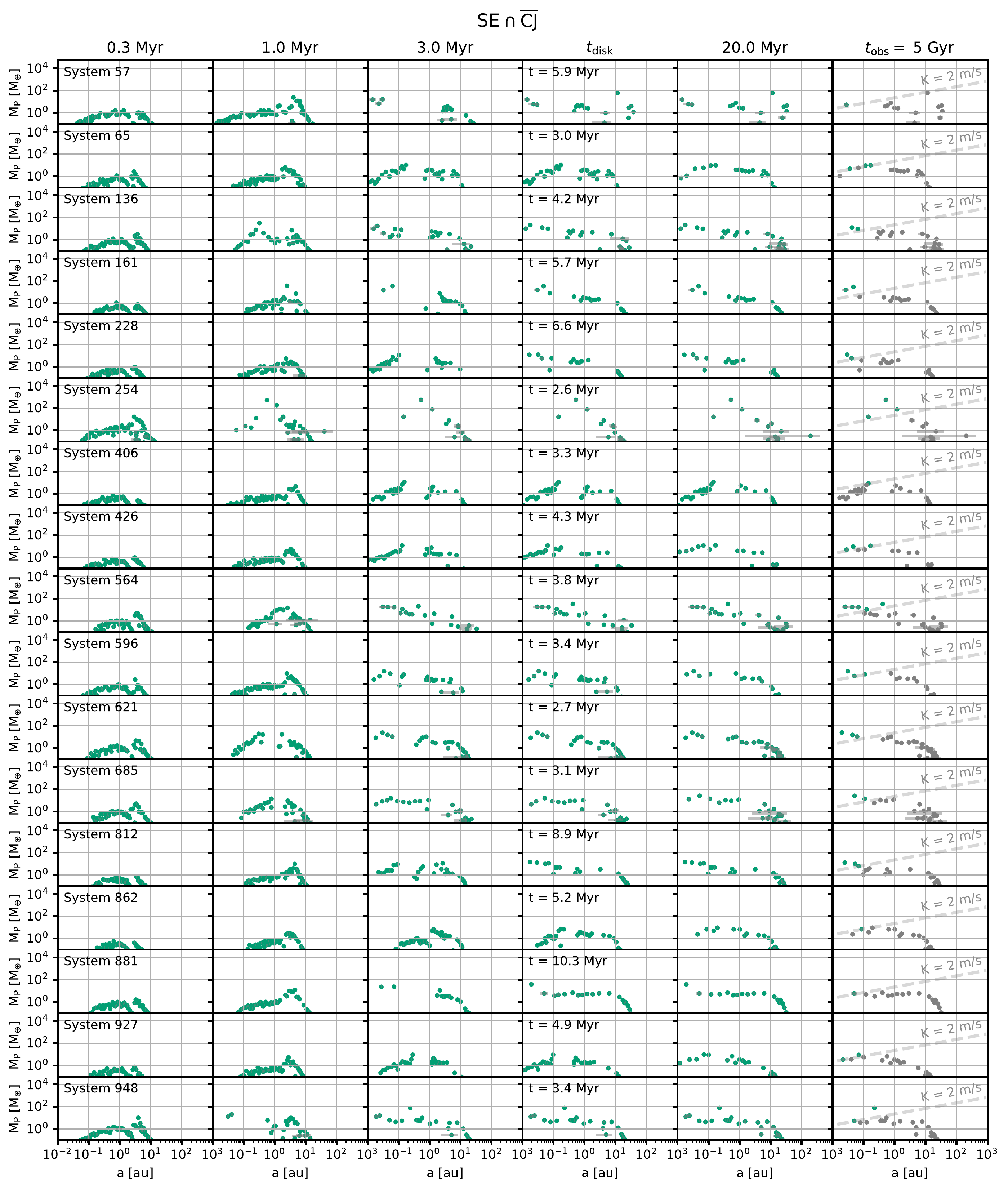}
        \caption{Same as Fig.~\ref{fig:sysEvo_SEandCJ}, but for systems containing super-Earths and no cold Jupiters.
        }
        \label{fig:sysEvo_SEandnoCJ}
\end{figure*}

\begin{figure*}
        \centering
        \includegraphics[width=\hsize]{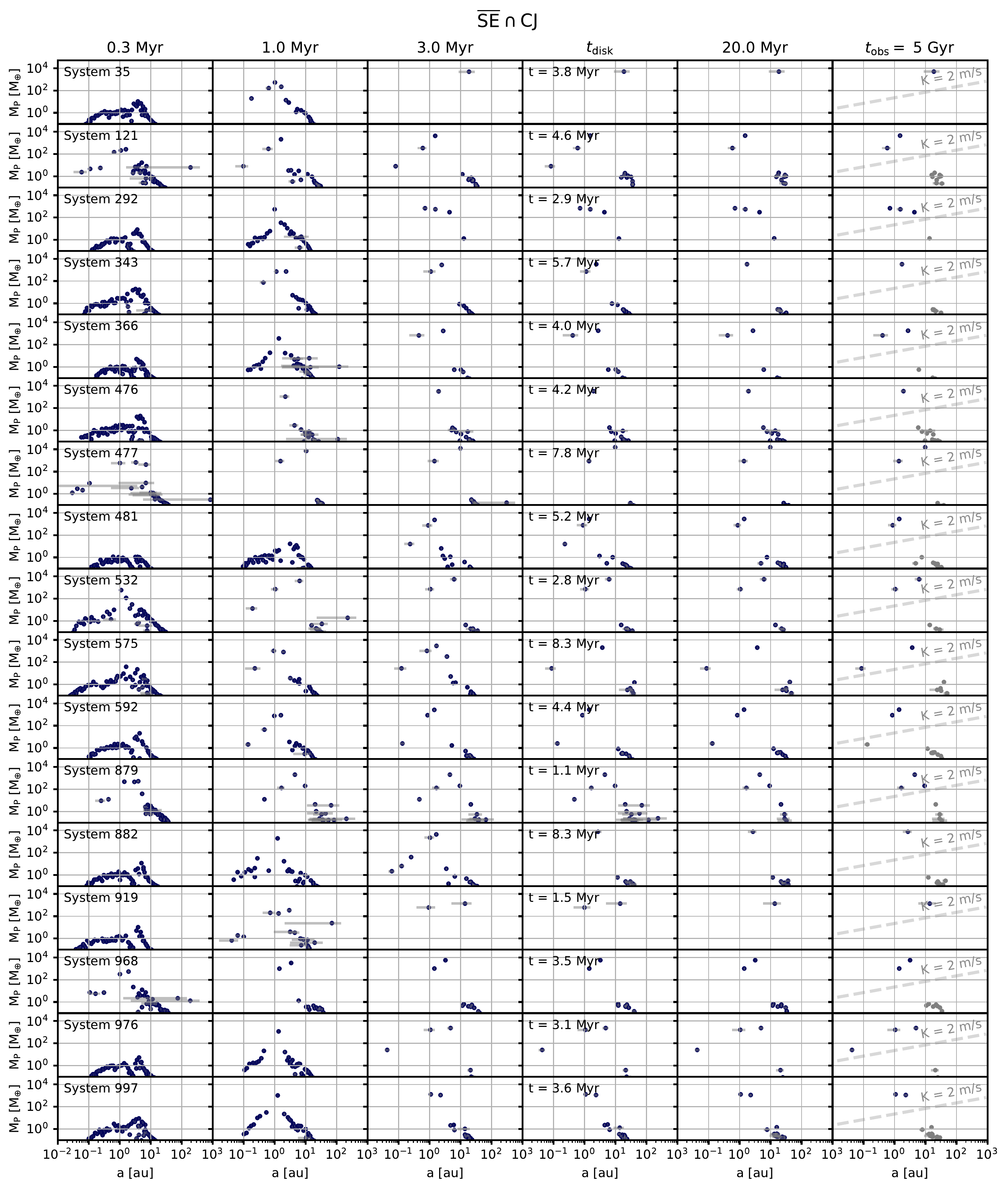}
        \caption{Same as Fig.~\ref{fig:sysEvo_SEandCJ}, but for systems containing cold Jupiters and no super-Earths.
        }
        \label{fig:sysEvo_noSEandCJ}
\end{figure*}

\end{appendix}

%\showthe\columnwidth

\end{document}